\documentclass[prd,twocolumn,showpacs,floatfix,amsmath,amssymb,nofootinbib]{revtex4}
\usepackage{bm}
\usepackage{graphicx}
\newcommand{\ud}{\mathrm{d}} 
\newcommand{\ui}{\mathrm{i}}

\newcommand{\GF}{G_\mathrm{F}}
\newcommand{\Hb}{\mathbf{H}}

\newcommand{\Sb}{\mathbf{S}}
\newcommand{\sB}{\mathbf{s}}
\newcommand{\eff}{\mathrm{eff}}
\newcommand{\HV}{\Hb_\mathrm{V}}
\newcommand{\muV}[1]{\mu_{\mathrm{V} #1}}
\newcommand{\HA}{\Hb_e}

\newcommand{\wsync}{\omega_\mathrm{sync}}
\newcommand{\basef}[1]{\hat{\mathbf{e}}^\mathrm{f}_{#1}}
\newcommand{\basem}[1]{\hat{\mathbf{e}}^\mathrm{m}_{#1}}
\newcommand{\basev}[1]{\hat{\mathbf{e}}^\mathrm{v}_{#1}}
\newcommand{\thetav}{\theta_\mathrm{v}}

\newcommand{\myfigsep}{0.04 \textwidth}
\newcommand{\myfigwid}{0.48 \textwidth}

\begin{document}
\title{Collective Neutrino Flavor Transformation In Supernovae}
\author{Huaiyu Duan}
\email{hduan@ucsd.edu}
\author{George M. Fuller}
\email{gfuller@ucsd.edu}
\affiliation{Department of Physics, %
University of California, San Diego, %
La Jolla, CA 92093-0319}
\author{Yong-Zhong Qian}
\email{qian@physics.umn.edu}
\affiliation{School of Physics and Astronomy, %
University of Minnesota, Minneapolis, MN 55455}
\date{\today}

\begin{abstract}
We examine coherent active-active channel neutrino flavor evolution
in environments where neutrino-neutrino forward scattering
can engender large-scale collective flavor transformation.
We introduce the concept of neutrino flavor isospin which
treats neutrinos and antineutrinos on an equal footing, and
which facilitates the analysis of neutrino systems in terms
of the spin precession analogy. We point out a key quantity,
the ``total effective energy'', which is conserved in several
important regimes. Using this concept, we analyze collective
neutrino and antineutrino flavor oscillation in the ``synchronized''
mode and what we term the ``bi-polar'' mode. We thereby are able
to explain why large collective flavor mixing can develop on
short timescales even when vacuum mixing angles are small in,
\textit{e.g.}, a dense gas of initially pure $\nu_e$ and $\bar\nu_e$
with an inverted neutrino mass hierarchy (an example of bi-polar oscillation).
In the context of the spin precession analogy, we find that the
co-rotating frame provides insights into more general systems,
where either the synchronized or bi-polar mode could arise.
For example, we use the co-rotating frame to demonstrate
how large flavor mixing in the bi-polar mode can occur in the
presence of a large and dominant matter background. We use
the adiabatic condition to derive a simple criterion for determining
whether the synchronized or bi-polar mode will occur.
Based on this criterion we predict that
neutrinos and antineutrinos emitted from a proto-neutron star
in a core-collapse supernova event can experience
synchronized and bi-polar flavor transformations in sequence
before conventional Mikhyev-Smirnov-Wolfenstein 
flavor evolution
takes over. This certainly will affect 
the analyses of future
supernova neutrino signals, and might affect the treatment
of shock re-heating
rates and nucleosynthesis depending on the depth at which
 collective transformation arises.
\end{abstract}


\maketitle
                                                              
\section{Introduction\label{sec:introduction}}

In both the early universe and in core-collapse supernovae,
neutrinos and antineutrinos can dominate energetics and can be 
instrumental in setting compositions (\textit{i.e.}, the neutron-to-proton
ratio). However, the way these particles couple to matter in these
environments frequently is flavor specific. Whenever there
are differences in the number fluxes or energy distribution
functions among the active neutrino species
($\nu_e$, $\bar\nu_e$, $\nu_\mu$, $\bar\nu_\mu$, $\nu_\tau$ and $\bar\nu_\tau$),
flavor mixing and conversion can be important
\cite{Fuller:1987aa,Fuller:1992aa,Sigl:1992fn,Qian:1993dg,Qian:1994wh,%
Pastor:2001iu,Schirato:2002tg,Balantekin:2004ug}.

In turn, the flavor conversion process becomes complicated and
nonlinear in environments with large effective neutrino and/or
antineutrino number densities
\cite{Fuller:1987aa,Pantaleone:1992xh,Sigl:1992fn,Samuel:1993uw,Qian:1994wh,%
Qian:1995ua}. 
In these circumstances
neutrino-neutrino forward scattering can become an important
determinant of the way in which neutrinos and antineutrinos
oscillate among flavor states. 

Two of the three vacuum mixing angles for the
active neutrinos are now measured. The third angle ($\theta_{13}$) is
constrained by experiments and is limited to values such that
$\sin^22\theta_{13}\lesssim 0.1$
(see, \textit{e.g.}, Ref.~\cite{Fogli:2005cq} for a review). 
In addition, the differences
of the squares of the neutrino mass eigenvalues 
are now measured, though the absolute masses and, therefore,
the neutrino mass hierarchy  remains unknown.

Both the solar and atmospheric neutrino mass-squared differences
are small, so small in fact that conventional matter-driven
Mikhyev-Smirnov-Wolfenstein (MSW) evolution
\cite{Wolfenstein:1977ue,Wolfenstein:1979ni,Mikheyev:1985aa}
would suggest that
neutrino and/or antineutrino flavor conversion occurs only far
out in the supernova envelope. On the other hand, it has been shown that
plausible conditions of neutrino flux in both the early shock
re-heating epoch and the later neutrino-driven wind,
$r$-process epoch, could provide the necessary condition
for neutrino-neutrino forward scattering induced large-scale flavor
conversion deep in the supernova environment \cite{Fuller:2005ae}.

The treatment of the flavor evolution of supernova neutrinos
remains a complicated problem, and the exact solution to this problem
may only be revealed by full self-consistent numerical simulations.
However, physical insights still can be gained by studying
somewhat simplified models of the realistic environments.
For example, one source of complication is that there are three active flavors
of neutrinos in play.
As the measured vacuum mass-squared difference for atmospheric neutrino 
oscillations ($\delta m_\mathrm{atm}^2\simeq 3\times 10^{-3}\ \mathrm{eV}^2$)
is much larger than that for solar neutrino oscillations
($\delta m_\odot^2\simeq 8\times 10^{-5}\ \mathrm{eV}^2$), the general problem of
three-neutrino mixing in many cases may be reduced to 
two separate cases of two-neutrino
mixing, each involving $\nu_e$ ($\bar\nu_e$) and some linear combination
of $\nu_\mu$ and $\nu_\tau$ ($\bar\nu_\mu$ and $\bar\nu_\tau$).
This reduction allows the possibility of visualizing
the neutrino flavor transformation as the rotation of 
a ``polarization vector'' in a three dimensional flavor space
\cite{Mikheyev:1986tj}. 
Different notations have been
developed around this concept (see, \textit{e.g.}, 
Refs.~\cite{Sigl:1992fn,Kostelecky:1993ys}). However, none
of these notations fully exhibits the symmetry
of particles and anti-particles in the SU(2) group
that governs the $2\times 2$ flavor transformation.

The equations of motion (e.o.m.) of a neutrino
``polarization vector'' is similar to those of a magnetic spin
precessing around magnetic fields. One naturally expects that
some collective behaviors may exist in dense neutrino gases
just as for magnetic spins in crystals. Indeed, it was observed in
numerical simulations that  neutrinos with different energies
in a dense gas act as if they have the same vacuum oscillation
frequency \cite{Samuel:1993uw}. This collective behavior
was later explained by drawing analogy to atomic spin-orbit
coupling in external fields and termed ``synchronization'' 
\cite{Pastor:2001iu}. 

Another, more puzzling, type of collective flavor transformation,
the ``bi-polar'' mode,
 has  been observed in the numerical simulations of a dense gas
of initially pure $\nu_e$ and $\bar\nu_e$  \cite{Kostelecky:1993dm}.
This type of collective flavor transformation
usually occurs on timescales much shorter than those of  
vacuum oscillations. Although the analytical solutions
to some simple examples of ``bi-polar'' systems
have been found \cite{Kostelecky:1994dt,Samuel:1995ri},
many aspects of these bi-polar systems still remain to be understood.
In particular, it seems counter-intuitive that, 
even for a small mixing angle, large flavor
mixing occurs in both the neutrino and antineutrino sectors
in a dense gas initially consisting of pure $\nu_e$ and $\bar\nu_e$
for an inverted mass hierarchy.

Both synchronized and bi-polar flavor transformation were
discovered in the numerical simulations aimed at
the early universe environment. It has been shown that
synchronized oscillation can also occur in
the supernova environment \cite{Pastor:2002we}. However, it
is not clear if supernova neutrinos can also have bi-polar
flavor transformation. If supernova neutrinos can have
collective synchronized and/or bi-polar oscillations, the questions
are then where these collective oscillations would occur
and how neutrino energy spectra would be modified.

In this paper we  try to answer the above questions.
In Sec.~\ref{sec:general-formalism} 
we will give the general equations governing the mixing of two neutrino 
flavors in the frequently used forms and introduce the notation
of neutrino flavor isospin, which treats neutrinos and antineutrinos
on an equal footing. We will also point out a key quantity,
the ``total effective energy'',
in analogy to the total energy of magnetic spin systems, which is
conserved in some interesting cases.
In Sec.~\ref{sec:synchronization} and Sec.~\ref{sec:bi-polar} 
we will analyze the synchronized and bi-polar neutrino systems
using the same framework in each case.
We will first describe and explain the
main features of these collective modes using the concept of
total effective energy. We then generalize these analyses 
by employing ``co-rotating frames''. We will
derive the criteria for the occurrence of these collective modes, and
discuss  the effects of an ordinary
matter background.
In Sec.~\ref{sec:supernovae} we will outline the
regions in supernovae where the neutrino mixing is
dominated by the synchronized, bi-polar and conventional MSW
flavor transformations. We will also describe
the typical neutrino mixing scenarios expected with different
matter density profiles.
In Sec.~\ref{sec:conclusions} we will summarize our new findings and
give our conclusions.

\section{General Equations Governing Neutrino Flavor Transformation%
\label{sec:general-formalism}}
We consider the mixing of two neutrino flavor eigenstates, 
say $|\nu_e\rangle$ and 
$|\nu_\tau\rangle$, which are linear combinations of the 
vacuum mass eigenstates 
$|\nu_1\rangle$ and $|\nu_2\rangle$ with eigenvalues
 $m_1$ and $m_2$, respectively:
\begin{subequations}
\begin{eqnarray}
|\nu_e\rangle&=&\cos\thetav|\nu_1\rangle+\sin\thetav|\nu_2\rangle,\\
|\nu_\tau\rangle&=&-\sin\thetav|\nu_1\rangle+\cos\thetav|\nu_2\rangle,
\end{eqnarray}
\end{subequations}
where $\thetav$ is the vacuum mixing angle. We take $\thetav<\pi/4$ and
refer to $\delta m^2\equiv m_2^2-m_1^2>0$ as the normal mass 
hierarchy and $\delta m^2<0$ as the
inverted mass hierarchy. When a neutrino
with energy $E_\nu$ propagates in matter, the evolution of its wavefunction 
in the flavor basis
\begin{equation}
\psi_{\nu}\equiv\left(\begin{array}{c}
a_{\nu_e}\\
a_{\nu_\tau}\end{array}\right)
\end{equation}
is governed by a Schr\"odinger-like equation
\begin{equation}
\ui\frac{\ud}{\ud t}\psi_{\nu}=({\cal{H}}_{\mathrm{v}}+{\cal{H}}_e)\psi_{\nu},
\label{eq:schrodinger}
\end{equation}
where $a_{\nu_e}$ and $a_{\nu_\tau}$ are the amplitudes for the neutrino to be in 
$|\nu_e\rangle$ and $|\nu_\tau\rangle$ at time $t$, respectively.
(This equation is ``Schr\"odinger-like'' because, unlike
the Schr\"odinger equation, we are here concerned with flavor evolution
at fixed energy and with relativistic leptons.)
The vacuum mass contribution ${\cal{H}}_{\mathrm{v}}$ 
to the propagation Hamiltonian in the flavor basis is
\begin{equation}
{\cal{H}}_{\mathrm{v}}=\frac{\delta m^2}{4E_\nu}\left(
\begin{array}{lr}
-\cos2\thetav&\sin2\thetav\\
\sin2\thetav&\cos2\thetav
\end{array}\right),
\label{eq:Hamiltonian}
\end{equation}
 and the contribution ${\cal{H}}_e$
due to forward scattering on electrons in the same basis is
\begin{equation}
{\cal{H}}_e=\frac{A}{2}\left(
\begin{array}{lr}
1&0\\
0&-1
\end{array}\right),
\end{equation}
where $A=\sqrt{2}\GF n_e$ with $n_e$ being the net electron number density. 
Eq.~\eqref{eq:schrodinger} also applies to the antineutrino wavefunction
\begin{equation}
\psi_{\bar\nu}\equiv\left(\begin{array}{c}
a_{\bar\nu_e}\\
a_{\bar\nu_\tau}\end{array}\right)
\end{equation}
if $A$ in ${\cal{H}}_e$ is replaced by $-A$.

When a large number of neutrinos and antineutrinos propagate through the same region 
of matter, their forward scattering on each other makes another contribution to the 
propagation Hamiltonian for each particle. For the $i$th neutrino, this contribution is 
\cite{Fuller:1987aa,Notzold:1987ik,Pantaleone:1992xh,Sigl:1992fn,Qian:1994wh}
\vspace{0.001in}
\begin{equation}
{\cal{H}}_{\nu\nu,i}=\frac{1}{2}\left(
\begin{array}{lr}
B&B_{e\tau}\\
B_{e\tau}^*&-B
\end{array}\right),
\end{equation}
where
\begin{widetext}
\begin{subequations}
\begin{eqnarray}
B&=&\sqrt{2}\GF\sum_j(1-\cos\Theta_{ij})\{n_{\nu,j}[(\rho_{\nu,j})_{ee}-
(\rho_{\nu,j})_{\tau\tau}]-n_{\bar\nu,j}[(\rho_{\bar\nu,j})_{ee}-
(\rho_{\bar\nu,j})_{\tau\tau}]\},\\
B_{e\tau}&=&2\sqrt{2}\GF\sum_j(1-\cos\Theta_{ij})[n_{\nu,j}(\rho_{\nu,j})_{e\tau}-
n_{\bar\nu,j}(\rho_{\bar\nu,j})_{e\tau}].
\end{eqnarray}
\end{subequations}
\end{widetext}
In the above equations, $\Theta_{ij}$ is the angle between 
the propagation directions
of the $i$th neutrino and the $j$th neutrino or antineutrino, 
and $n_{\nu,j}$ ($n_{\bar\nu,j}$) 
and $\rho_{\nu,j}$ ($\rho_{\bar\nu,j}$) are the number density and 
single-particle 
flavor-basis density matrix of the $j$th neutrino (antineutrino), 
respectively. Specifically, 
\begin{subequations}
\begin{align}
\rho_\nu&=\left(\begin{array}{lr}
|a_{\nu_e}|^2&a_{\nu_e}a_{\nu_\tau}^*\\
a_{\nu_e}^*a_{\nu_\tau}&|a_{\nu_\tau}|^2
\end{array}\right),
\label{eq:rhonu} \\
\intertext{and}
\rho_{\bar\nu}&=\left(\begin{array}{lr}
|a_{\bar\nu_e}|^2&a_{\bar\nu_e}^*a_{\bar\nu_\tau}\\
a_{\bar\nu_e}a_{\bar\nu_\tau}^*&|a_{\bar\nu_\tau}|^2
\end{array}\right),
\label{eq:rhonubar}
\end{align}
\end{subequations}
where we have adopted the convention for the density matrix of
an antineutrino in Ref.~\cite{Sigl:1992fn}.
The neutrino-neutrino forward scattering contribution for an antineutrino can be
obtained by making the substitution $B\to-B$ and $B_{e\tau}\to-B_{e\tau}^*$ in
${\cal{H}}_{\nu\nu,i}$.

The single-particle density matrices in Eqs.~\eqref{eq:rhonu} and \eqref{eq:rhonubar}
can be written in the form
\begin{equation}
\rho=\frac{1}{2}\left(1+\mathbf{P}\cdot\bm{\sigma}\right),
\end{equation}
where $\mathbf{P}$ is the polarization vector in the three-dimensional (Euclidean)
flavor space
and $\bm{\sigma}$ represents the Pauli matrices. Explicitly, the polarization vectors
in column form are 
\begin{equation}
\mathbf{P}_\nu=\left(\begin{array}{c}
2\mathrm{Re}(a_{\nu_e}^*a_{\nu_\tau})\\
2\mathrm{Im}(a_{\nu_e}^*a_{\nu_\tau})\\
|a_{\nu_e}|^2-|a_{\nu_\tau}|^2
\end{array}\right),
\label{eq:pnu}
\end{equation}
and
\begin{equation}
\mathbf{P}_{\bar\nu}=\left(\begin{array}{c}
2\mathrm{Re}(a_{\bar\nu_e}a_{\bar\nu_\tau}^*)\\
2\mathrm{Im}(a_{\bar\nu_e}a_{\bar\nu_\tau}^*)\\
|a_{\bar\nu_e}|^2-|a_{\bar\nu_\tau}|^2
\end{array}\right).
\label{eq:pnubar}
\end{equation}
By straightforward algebra, it can be shown that the Schr\"odinger-like 
equation
\begin{equation}
\ui\frac{\ud}{\ud t}\psi_{\nu,i}=({\cal{H}}_{\mathrm{v}}+{\cal{H}}_e+
{\cal{H}}_{\nu\nu,i})\psi_{\nu,i}
\end{equation}
and a similar equation for an antineutrino lead to  \cite{Sigl:1992fn}
\begin{widetext}
\begin{subequations}
\begin{eqnarray}
\frac{\ud}{\ud t}\mathbf{P}_{\nu,i}&=&\mathbf{P}_{\nu,i}\times\left[
\frac{\delta m^2}{2E_{\nu,i}}\left(\begin{array}{c}
-\sin2\thetav\\
0\\
\cos2\thetav\end{array}\right)
-\sqrt{2}\GF n_e\left(\begin{array}{c}
0\\
0\\
1\end{array}\right)\right.
\left.-\sqrt{2}\GF\sum_j(1-\cos\Theta_{ij})(n_{\nu,j}\mathbf{P}_{\nu,j}
-n_{\bar\nu,j}\mathbf{P}_{\bar\nu,j})\right],
\label{eq:pnuev}\\
\frac{\ud}{\ud t}\mathbf{P}_{\bar\nu,i}&=&\mathbf{P}_{\bar\nu,i}\times\left[
-\frac{\delta m^2}{2E_{\bar\nu,i}}\left(\begin{array}{c}
-\sin2\thetav\\
0\\
\cos2\thetav\end{array}\right)
-\sqrt{2}\GF n_e\left(\begin{array}{c}
0\\
0\\
1\end{array}\right)\right.
\left.-\sqrt{2}\GF\sum_j(1-\cos\Theta_{ij})(n_{\nu,j}\mathbf{P}_{\nu,j}
-n_{\bar\nu,j}\mathbf{P}_{\bar\nu,j})\right].
\label{eq:pnubarev}
\end{eqnarray}
\end{subequations}
\end{widetext}

The three real components of the polarization vector 
contain the same information 
as the two complex amplitudes of the wavefunction
except for an overall phase which is irrelevant
for flavor transformation. 
Therefore, Eqs.~\eqref{eq:pnuev}
and \eqref{eq:pnubarev} are equivalent to the Schr\"odinger-like equations. 
Eqs.~\eqref{eq:pnuev} and \eqref{eq:pnubarev} appear to 
suggest a geometric picture 
of precessing polarization vectors. This picture has 
been discussed quite extensively 
in the literature (see, \textit{e.g.}, 
\cite{Kim:1987ss,Kim:1987bv,Pastor:2001iu}) 
and shown to be especially helpful 
in understanding flavor transformation 
when neutrino self-interaction (\textit{i.e.}, neutrino-neutrino 
forward scattering) is important.
To facilitate the use of this picture, we briefly discuss 
the physics behind it and introduce 
some notations. 

For simplicity, we first consider only the contributions 
${\cal{H}}_{\mathrm{v}}$ and ${\cal{H}}_e$ to the propagation Hamiltonian $\cal{H}$
for a neutrino. In this case, we can write
\begin{equation}
{\cal{H}}={\cal{H}}_\mathrm{v}+{\cal{H}}_e=-\frac{\bm{\sigma}}{2}\cdot(\muV{}\HV+\HA),
\label{eq:hve}
\end{equation}
where 
\begin{eqnarray}
\muV{}&\equiv&\frac{\delta m^2}{2E_\nu},\\
\HV&\equiv&-\basef{x}\sin2\thetav+\basef{z}\cos2\thetav,\\
\HA&\equiv&-\basef{z}\sqrt{2}\GF n_e,
\end{eqnarray}
with $\basef{x}$ and $\basef{z}$ being the unit vectors in the $x$- and $z$-directions
in the flavor basis, respectively. Eq.~\eqref{eq:hve} takes the form of the interaction 
between the ``magnetic moment'' $\bm{\mu}=\gamma\sB$ of a spin-$\frac{1}{2}$ 
particle and an external ``magnetic field'' $\Hb=\Hb^\eff/\gamma$ with
\begin{equation}
\Hb^\eff\equiv\muV{}\HV+\HA.
\end{equation}
Here $\gamma$ is the ``gyromagnetic ratio'' and can be chosen arbitrarily.
Classically, the spin $\sB$ would experience a torque 
$\bm{\tau}=\bm{\mu}\times\Hb=\sB\times\Hb^\eff$ and its e.o.m.~would 
be given by the angular momentum theorem:
\begin{equation}
\frac{\ud}{\ud t}\sB=\bm{\tau}=\sB\times\Hb^\eff.
\label{eq:classeom}
\end{equation}
By Ehrenfest's theorem, the quantum mechanical description of a system
has the same form as the classical e.o.m.~provided that all physical
observables are replaced by the expectation values of their quantum mechanical
operators. In the present case, if we replace $\sB$ in Eq.~\eqref{eq:classeom} by
\begin{equation}
\sB_\nu\equiv\psi_\nu^\dagger\frac{\bm{\sigma}}{2}\psi_\nu=\frac{\mathbf{P}_\nu}{2},
\label{eq:nfis}
\end{equation}
then neutrino flavor transformation governed by
${\cal{H}}={\cal{H}}_\mathrm{v}+{\cal{H}}_e$
can be described quantum mechanically by
\begin{equation}
\frac{\ud}{\ud t}\sB_\nu=\sB_\nu\times\Hb^\eff,
\label{eq:prec}
\end{equation}
which is the same as Eq.~\eqref{eq:pnuev} in the absence of 
neutrino self-interaction. 
Clearly, the operator $\bm{\sigma}/2$ in 
Eqs.~\eqref{eq:hve} and \eqref{eq:nfis} represents a 
fictitious spin in the neutrino flavor space, which may 
be appropriately called
the neutrino flavor isospin (NFIS). The flavor eigenstates $|\nu_e\rangle$ and 
$|\nu_\tau\rangle$ correspond to the up and down eigenstates, respectively, of 
the $z$-component of $\bm{\sigma}/2$. We will loosely refer to the expectation 
value $\sB_\nu$ of this operator as the NFIS and use it 
instead of the polarization 
vector $\mathbf{P}_\nu$ to describe neutrino 
flavor transformation. 
The $z$-component of a NFIS $\sB_\nu$ is of
special importance  as it  determines the
probability for the corresponding neutrino to be in $|\nu_e\rangle$:
\begin{equation}
s_{\nu z}^\mathrm{f}\equiv \sB_\nu\cdot\basef{z} 
=\frac{|a_{\nu_e}|^2-|a_{\nu_\tau}|^2}{2}= |a_{\nu_e}|^2-\frac{1}{2}.
\end{equation}
Therefore, for a neutrino, $s_{\nu z}^\mathrm{f}=1/2$, $-1/2$ and 0 correspond to
$|\nu_e\rangle$, $|\nu_\tau\rangle$ and a maximally  mixed state, respectively.

Adiabatic MSW flavor conversion
  has a simple explanation in this ``magnetic spin'' analogy.
For illustrative purposes we assume $\delta m^2>0$ and $\thetav\ll 1$.
As a $\nu_e$ propagates from a region with large matter density,
\textit{e.g.}, the core of the sun, to a region of very little ordinary matter,
$\Hb^\eff$ changes its direction from $\sim -\basef{z}$ to $\HV\sim\basef{z}$.
If the density of electrons $n_e$ changes only slowly along the way
(adiabatic process),
$\Hb^\eff$ also changes slowly, and the NFIS $\sB_\nu$ corresponding to the
neutrino is always anti-aligned with $\Hb^\eff$. 
Therefore, the neutrino
originally in the $\nu_e$ eigenstate ($\sB_\nu=\basef{z}/2$)
is now mostly in the $\nu_\tau$ eigenstate ($\sB_\nu\simeq-\basef{z}/2$).

It is useful to illustrate the criterion for adiabadicity
of this process in the ``magnetic spin'' analogy.
First, we note that the probabilities for a neutrino to be in instantaneous 
mass eigenstates $\nu_{\mathrm{L}}$ (light) and $\nu_{\mathrm{H}}$ (heavy) are
\begin{subequations}
\begin{eqnarray}
|a_{\nu_\mathrm{L}}|^2 &=& \frac{1}{2} + \sB_\nu\cdot\basem{z}
= \frac{1+\cos2\theta}{2},\\
|a_{\nu_\mathrm{H}}|^2 &=& \frac{1}{2} - \sB_\nu\cdot\basem{z}
= \frac{1-\cos2\theta}{2},
\end{eqnarray}
\end{subequations}
respectively, where $2\theta$ is the angle between the directions of
$\sB_\nu$ and $\basem{z}\equiv\Hb^\eff/|\Hb^\eff|$, and
$\basem{i}$ are the unit vectors for the instantaneous mass basis.
In the MSW picture, $\theta$ is the instantaneous matter mixing angle.
In an adiabatic process,
$|a_{\nu_\mathrm{L}}|^2$ and $|a_{\nu_\mathrm{H}}|^2$ are constant, and
so is $\theta$. Using Eq.~\eqref{eq:prec} we have
\begin{equation}
\frac{1}{2}
\frac{\ud}{\ud t} (\cos2\theta) = \frac{\ud}{\ud t} (\sB_\nu\cdot\basem{z})
=\sB_\nu\cdot\frac{\ud}{\ud t}\basem{z}.
\label{eq:dcos2theta}
\end{equation}
On a timescale 
$\delta t \gtrsim 2\pi/|\Hb^\eff|$, 
$\sB_\nu$ has rotated by at least one cycle around $\Hb^\eff$.
If $\Hb^\eff$ changes its direction only by
a small angle $\delta\phi\equiv|\ud \basem{z}/\ud t|\delta t \ll 2\pi$ 
during $\delta t$, then
$\sB_\nu$ in Eq.~\eqref{eq:dcos2theta} averages to 
$(\sB_\nu\cdot\basem{z})\basem{z}$.
Noting that 
$\basem{z}\cdot(\ud\basem{z}/\ud t)=(1/2)\ud(|\basem{z}|^2)/\ud t=0$,
one can see that the angle $\theta$ is unchanged in this process.
Therefore, the criterion for a MSW flavor transformation to be adiabatic is
\begin{equation}
\left|\frac{\ud}{\ud t}\basem{z}\right|=
\frac{|\dot\Hb^\eff\times\Hb^\eff|}{|\Hb^\eff|^2} \ll |\Hb^\eff|,
\label{eq:adiabatic-cond}
\end{equation}
which is equivalent to saying that the rate of change of 
the direction of the ``magnetic field''
$\Hb^\eff$ is much smaller than the rotating rate of the``magnetic spin''
$\sB_\nu$ around $\Hb^\eff$.

The full version of Eq.~\eqref{eq:pnuev} can be obtained by extending 
the Hamiltonian in Eq.~\eqref{eq:Hamiltonian} to include 
\begin{equation}
{\cal{H}}_{\nu\nu,i}=-\frac{\bm{\sigma}}{2}\cdot\sum_j\mu_{ij}
\left(n_{\nu,j}\frac{\mathbf{P}_{\nu,j}}{2}-n_{\bar\nu,j}
\frac{\mathbf{P}_{\bar\nu,j}}{2}\right),
\label{eq:hnu}
\end{equation}
where
\begin{equation}
\mu_{ij}\equiv-2\sqrt{2}\GF(1-\cos\Theta_{ij}).
\label{eq:mu-ij}
\end{equation}
We define the NFIS for an antineutrino as%
\footnote{The two fundamental representations $\mathbf{2}$ and 
$\overline{\mathbf{2}}$
of the SU(2) group generated by the Pauli matrices
are equivalent. These representations are related to each other 
by the transformation $\sigma_y$, and 
$\tilde\psi_{\bar\nu}\equiv\sigma_y\psi_{\bar\nu}$ transforms
in exactly the same way as does $\psi_\nu$ under rotation.
Defining $\sB_{\bar\nu}\equiv \tilde\psi_{\bar\nu}^\dagger(\bm{\sigma}/2)
\tilde\psi_{\bar\nu}$, one naturally  obtains the minus 
sign in Eq.~\eqref{eq:nfis-anu}.}
\begin{equation}
\sB_{\bar\nu}\equiv-\frac{\mathbf{P}_{\bar\nu}}{2},
\label{eq:nfis-anu}
\end{equation}
so that the terms related to neutrinos and 
antineutrinos appear symmetrically in ${\cal{H}}_{\nu\nu,i}$.
The probability for an antineutrino to be in $|\bar\nu_e\rangle$
is determined from
\begin{equation}
s_{\bar\nu z}^\mathrm{f}\equiv \sB_{\bar\nu}\cdot\basef{z} 
=\frac{|a_{\bar\nu_\tau}|-|a_{\bar\nu_e}|^2}{2}= \frac{1}{2}-|a_{\bar\nu_e}|^2.
\end{equation}
For an antineutrino, $s_{\bar\nu z}^\mathrm{f}=1/2$, $-1/2$ and 0 correspond
to $|\bar\nu_\tau\rangle$, $|\bar\nu_e\rangle$ and 
a maximally mixed state, respectively.

Now Eqs.~\eqref{eq:pnuev} and \eqref{eq:pnubarev} can be rewritten in terms 
of the NFIS's in a more compact way
\begin{equation}
\frac{\ud}{\ud t}\sB_i=\sB_i\times\left(\muV{,i}\HV+\HA
+\sum_j\mu_{ij}n_{\nu,j}\sB_j\right),
\label{eq:nfisev}
\end{equation}
with the understanding that 
\begin{equation}
\muV{,i}\equiv\left\{\begin{array}{ll}
\delta m^2/(2E_{\nu,i}) &\text{for a neutrino},\\
-\delta m^2/(2E_{\bar\nu,i})& \text{for an antineutrino},\end{array}\right.
\label{eq:muV-def}
\end{equation}
and that the sum runs over both neutrinos and antineutrinos.
We also define a total effective energy (density) $\cal{E}$
for a system of neutrinos and
antineutrinos that interact with a matter background as well as among themselves
through forward scattering:
\begin{equation}
{\cal{E}}\equiv-\sum_in_{\nu,i}\sB_i\cdot\Hb^\eff_i
-\frac{1}{2}\sum_{i,j}\mu_{ij}n_{\nu,i}n_{\nu,j}\sB_i\cdot\sB_j,
\end{equation}
where 
\begin{equation}
\Hb^\eff_i\equiv\muV{,i}\HV+\Hb_e.
\end{equation}
We note that  this effective energy should not be confused
 with the physical energies
of neutrinos and antineutrinos.
It can be shown from Eq.~\eqref{eq:nfisev} that $\mathcal{E}$
is constant 
if $n_e$ and all the $n_{\nu,i}$'s and $\mu_{ij}$'s are also constant.
The concept of the total effective energy will prove useful in understanding
collective flavor transformation in a dense neutrino gas.

In the early universe  the neutrino gas is isotropic, and
\begin{equation}
\mu_{ij} \rightarrow \mu_\nu \equiv -2\sqrt{2}\GF.
\end{equation}
For illustrative purposes we will assume this 
isotropy condition in most of what follows.
We will discuss the effects of the anisotropic  supernova
neutrino distributions in Sec.~\ref{sec:supernovae}.

\section{Synchronized Flavor Transformation%
\label{sec:synchronization}} 
In a dense neutrino gas NFIS's are coupled to each other through 
self-interaction and may exhibit collective behaviors. As
discovered in the numerical simulations of Ref.~\cite{Samuel:1993uw},
neutrinos with different energies in a dense gas act as if they are oscillating 
with the same frequency. This collective behavior was referred to as
``synchronized'' flavor oscillations in the literature
and explained in Ref.~\cite{Pastor:2001iu} 
by drawing analogy to atomic spin-orbit coupling in external magnetic fields.
In this section we will first review the characteristics of a simple 
synchronized NFIS system from the perspective of the conservation of
the total energy $\mathcal{E}$ of the NFIS system. We will then
extend the discussion to more general synchronized NFIS systems using the
concept of a ``co-rotating frame'' and
demonstrate the criteria for a NFIS system to be in the synchronized mode.
We will show that the stability of a synchronized system is secured
by the conservation of the total effective energy.
In the last part of the section we will look into the 
problem of synchronized flavor
transformation in the presence of ordinary matter, which is relevant
for the supernova environment.

\subsection{A Simple Example of Synchronized Flavor Transformation%
\label{sec:simple-synchronization}}
We start with a simple case of a uniform and isotropic neutrino 
gas with no matter background ($n_e=0$). The gas initially consists of pure 
neutrinos with a finite energy range corresponding to
$|\muV{,i}|\leq|\muV{,i}|_{\rm max}$, and
all the $n_{\nu,i}$'s stay constant.
The e.o.m.~of a single NFIS is
\begin{equation}
\frac{\ud}{\ud t}\sB_i=\sB_i\times(\muV{,i}\HV+\mu_\nu\Sb),
\label{eq:isosync}
\end{equation}
where
\begin{equation}
\Sb\equiv\sum_j n_{\nu,j}\sB_j
\end{equation}
is the total NFIS (density) of the gas. 
(The NFIS density for an individual ``spin'' $\sB_j$
is just $n_{\nu,j}\sB_j$.)
Summing Eq.~\eqref{eq:isosync}
over all neutrinos, we obtain
\begin{equation}
\frac{\ud}{\ud t}\Sb=\sum_i\muV{,i}n_{\nu,i}\sB_i\times\HV.
\label{eq:isonfis}
\end{equation}
Following the discussion at the end of the preceding section,
the evolution of the individual ($\sB_i$) and the total ($\Sb$) NFIS 
obeys conservation of the total effective energy
\begin{subequations}
\begin{align}
{\cal{E}}&=-\sum_i\muV{,i}n_{\nu,i}\sB_i\cdot\HV
\nonumber\\
&\quad-\frac{1}{2}\sum_{i,j}\mu_{ij}n_{\nu,i}n_{\nu,j}\sB_i\cdot\sB_j\\
&=-\sum_i\muV{,i}n_{\nu,i}\sB_i\cdot\HV-\frac{\mu_\nu}{2}\Sb^2.
\label{eq:energycons}
\end{align}
\end{subequations}

An interesting limit is
\begin{equation}
|\mu_\nu\Sb|\gg|\muV{,i}|_\mathrm{max}.
\label{eq:synclim}
\end{equation}
Noting that each $\sB_i$ has a magnitude of 1/2 and $\HV$ has a 
magnitude of unity, we see that
\begin{equation}
{\cal{E}}\simeq-\frac{\mu_\nu}{2}\Sb^2\simeq\mathrm{const.}
\label{eq:effenlim}
\end{equation}
in the above limit. Therefore, 
a gas with a large initial total NFIS $\Sb$ evolves in
such a way that it roughly maintains the
magnitude of its $\Sb$. For such a gas, Eq.~\eqref{eq:isosync} reduces to
\begin{equation}
\frac{\ud}{\ud t}\sB_i\simeq\mu_\nu\sB_i\times\Sb,
\label{eq:isosynclim}
\end{equation}
which means that each $\sB_i$ precesses around the total NFIS 
with a fixed common (angular) frequency
\begin{equation}
\omega_\nu\equiv|\mu_\nu\Sb|.
\end{equation}
Eq.~\eqref{eq:isonfis} shows that $\Sb$ evolves on a timescale 
$\gtrsim 2\pi/|\muV{,i}|_\mathrm{max}$. Consequently, over a period $\delta t$
satisfying
\begin{equation}
\frac{2\pi}{\omega_\nu}\ll\delta t\ll\frac{2\pi}{|\muV{,i}|_\mathrm{max}},
\end{equation}
$\sB_i$ averages out to be $(\sB_i\cdot\Sb)\Sb/\Sb^2$
and Eq.~\eqref{eq:isonfis} effectively becomes
\begin{equation}
\frac{\ud}{\ud t}\Sb\simeq\wsync\Sb\times\HV,
\label{eq:isonfislim}
\end{equation}
where
\begin{equation}
\wsync=\langle\muV{}\rangle
\equiv\sum_i\frac{\muV{,i}n_{\nu,i}\sB_i\cdot\Sb}{\Sb^2}.
\label{eq:wsync}
\end{equation}
It can be shown from Eqs.~\eqref{eq:isonfis}, \eqref{eq:effenlim}, and 
\eqref{eq:isosynclim} that $\wsync\simeq\mathrm{const.}$ 
Therefore, $\Sb$ precesses around $\HV$ with a fixed frequency
$\wsync$ while the individual $\sB_i$'s precess around $\Sb$ with a fixed
common frequency $\omega_\nu$. This collective behavior of a dense 
neutrino gas is usually referred to as synchronized flavor oscillations
\cite{Pastor:2001iu}.

\subsection{General Synchronized Systems%
\label{sec:general-synchronization}}

Synchronization can occur not only in dense neutrino gases but
also in dense antineutrino gases and gases including both neutrinos
and antineutrinos. Noting that the NFIS's for neutrinos and antineutrinos
essentially only differ by the signs in $\muV{,i}$'s
[see Eq.~\eqref{eq:muV-def}], one can
 repeat the same arguments
 in Sec.~\ref{sec:simple-synchronization} for these more generalized cases.
Instead of doing so, we want to proceed from a new perspective, which
demonstrates some of the benefits of the NFIS notation.

We consider a  reference frame rotating with an angular
velocity of $-\Omega\HV$. In this co-rotating frame,
Eqs.~\eqref{eq:isosync} and \eqref{eq:isonfis}
take the form
\begin{subequations}
\begin{eqnarray}
\dot{\tilde{\sB}}_i&=&\tilde{\sB}_i \times (\tilde{\mu}_{\mathrm{V},i}\HV 
+ \mu_\nu{\widetilde{\Sb}}),\\
\dot{\widetilde{\Sb}}&=&\sum_i\tilde{\mu}_{\mathrm{V},i}n_{\nu,i}\tilde{\sB}_i 
\times \HV,
\end{eqnarray}
\end{subequations}
where $\tilde{\sB}_i$ ($\widetilde{\Sb}$) and $\dot{\tilde{\sB}}_i$ ($\dot{\widetilde{\Sb}}$)
are $\sB_i$ ($\Sb$) and its time derivative 
in terms of their $x$-, $y$-, and $z$-components in the co-rotating frame, and
\begin{equation}
\tilde{\mu}_{\mathrm{V},i} \equiv \muV{,i} - \Omega.
\end{equation}
It is clear that one can set $\tilde{\mu}_{\mathrm{V}}$ of a NFIS 
to any value by choosing an appropriate co-rotating frame, and a
NFIS for an antineutrino in the lab frame becomes a neutrino
in some co-rotating frame. For example, the NFIS in the lab frame with
$\sB=-\basef{z}/2$ and $\muV{}=-\delta m^2/2E$ corresponds to
a $\bar\nu_e$ with energy $E$. In a co-rotating frame with
$\Omega=-\delta m^2/E$ the NFIS has $\tilde\sB=-\basef{z}/2$
and $\tilde{\mu}_{\mathrm{V}}=\delta m^2/2E$, which corresponds
to a $\nu_\tau$ with energy $E$. 
Therefore, the NFIS notation really treats neutrinos and
antineutrinos on an equal footing. 

Because $\tilde\Sb$ and $\Sb$ are the same
vector in two different frames, the synchronization of the NFIS's
in one frame means the synchronization in any frame. Consequently,
synchronization can occur in dense antineutrino gases and gases 
of both neutrinos and antineutrinos just as it can occur in
pure neutrino gases as long as Eq.~\eqref{eq:synclim} is satisfied
in some co-rotating frame.

As we have seen, $|\muV{,i}|_{\max}$ is not uniquely determined
and can have different values in different co-rotating frames.
However, we note that the relative spread of the
individual values of the $\muV{,i}$'s
of the NFIS's is an intrinsic property of a NFIS system
and is co-rotating frame invariant.
For a co-rotating frame with
\begin{equation}
\Omega=\frac{(\muV{,i})_{\rm min}+(\muV{,i})_{\rm max}}{2},
\end{equation}
one has
\begin{equation}
|\tilde{\mu}_{{\rm V},i}|_\mathrm{max}=\Delta\muV{},
\end{equation}
where
\begin{equation}
\Delta\muV{}\equiv \frac{(\muV{,i})_\mathrm{max} - (\muV{,i})_\mathrm{min}}{2}
\end{equation}
measures the spread of the $\muV{,i}$'s in the NFIS system.
Synchronization can be obtained if
\begin{equation}
|\mu_\nu\Sb|\gg \Delta\muV{}.
\label{eq:sync-cond-weak}
\end{equation}

When applying this condition
to astrophysical environments such as the early universe and supernovae,
we must consider the meaning of $\Delta\muV{}$ 
as neutrinos in these environments formally
have an infinite energy range. 
One interesting scenario is where the distribution of 
NFIS density as a function of $\muV{}$ has a single dominant peak.
An example  is the neutronization burst
in a core-collapse supernova event where the neutrinos emitted 
are dominantly $\nu_e$ with a Fermi-Dirac-like energy
distribution $f_{\nu_e}(E)$. 
For this case a natural estimate of $\Delta\muV{}$
is the half-width of the distribution function $f_{\nu_e}(\muV{})$, 
where $f_{\nu_e}(\muV{})$ is obtained from $f_{\nu_e}(E)$ using 
the relation $\muV{}(E)=\delta m^2/2E$.

Another interesting scenario is where the distribution of
NFIS density as a function of $\muV{}$   has two dominant peaks.
An example of this scenario is the Kelvin-Helmholtz cooling phase
of a proto-neutron star in a core-collapse supernova event
where the neutrinos emitted are mostly (in number) $\nu_e$ and $\bar\nu_e$.
For this scenario one can take 
\begin{equation}
\Delta\muV{}\simeq\left|\frac{\delta m^2}{2E_{\nu_e}}+
\frac{\delta m^2}{2E_{\bar\nu_e}}\right|,
\label{eq:nue-anue-sync-cond}
\end{equation}
where $E_{\nu_e}$ and $E_{\bar\nu_e}$ are the peak energies of
the $\nu_e$ and $\bar\nu_e$ energy spectra, respectively.

For more complicated scenarios,
the criterion to obtain synchronized flavor oscillations
can be compared to the criterion for an adiabatic MSW flavor conversion.
If a NFIS system has been tested to be in a synchronized mode
using the analyses in 
Sec.~\ref{sec:simple-synchronization} in some co-rotating frame,
each individual NFIS $\tilde{\sB}$ should precess around the 
total NFIS $\widetilde{\Sb}$ with a fixed angle. This is the same 
``tracking'' behavior as in the adiabatic MSW flavor transformation
process discussed in Sec.~\ref{sec:general-formalism} except that
$\widetilde{\Sb}$ now takes the place of $\Hb^\eff$ in 
Eq.~\eqref{eq:adiabatic-cond}.
Because $\widetilde{\Sb}$ slowly rotates around $\HV$ with angular frequency
$\tilde\omega_{\mathrm{sync}}=\langle\tilde\mu_{\mathrm{V}}\rangle$, 
the adiabatic criterion yields
\begin{equation}
|\mu_\nu\widetilde{\Sb}|
\gtrsim |\langle\tilde\mu_{\mathrm{V}}\rangle|\cdot|\sin\xi|,
\label{eq:necessary-sync-cond}
\end{equation}
where $\xi$ is the angle between the directions of $\widetilde{\Sb}$ and $\HV$.
Eq.~\eqref{eq:necessary-sync-cond} provides a necessary
condition for synchronization. 
Practically one may use
\begin{equation}
|\mu_\nu\Sb|\gtrsim |\langle\muV{}\rangle|
\label{eq:sync-cond}
\end{equation}
as the criterion for synchronization,
where $\langle\muV{}\rangle$ is evaluated using Eq.~\eqref{eq:wsync}
with all the relevant neutrino and antineutrino energy distributions. 

We now make some comments on the stability of the synchronized mode.
Because neutrinos with different energies 
have different vacuum oscillation frequencies,
one may think that the NFIS's will develop relative phases and that
the resulting 
destructive interference will break the synchronization, \textit{i.e.},
reducing $\Sb$ to approximately 0. 
Indeed, using Eq.~\eqref{eq:isonfis} one can see that
\begin{subequations}
\begin{align}
\frac{\ud }{\ud t} \Sb^2 &= 2\Sb\cdot\dot\Sb\\
&=\sum_{ij}n_{\nu,i}n_{\nu,j}(\muV{,i}-\muV{,j})(\sB_j\times\sB_i)\cdot\HV
\end{align}
\end{subequations}
is generally not zero, and therefore, $|\Sb|$ varies with time.
However, Eq.~\eqref{eq:isosync}, from which Eq.~\eqref{eq:isonfis}
is derived, can be used to show that the total effective energy $\mathcal{E}$
is conserved and the total NFIS $\Sb$ roughly maintains constant magnitude
if the $n_{\nu,i}$'s do not vary with time.
[see Eq.~\eqref{eq:effenlim}]. 
In this case, destructive interference stemming from the relative phases
of different NFIS's  cannot completely destroy synchronized flavor oscillations.
 On the other hand,
if $|\Sb|\simeq 0$ initially, no significant synchronization
of NFIS's can occur spontaneously. This result is in accord
with the lengthy study in Ref.~\cite{Pantaleone:1998xi}.

\subsection{Synchronized Flavor Transformation with a Matter Background%
\label{sec:synchronization-matter}}
We now discuss the effects of a matter background on 
synchronized flavor transformation in dense gases of
neutrinos and/or antineutrinos.
The relevant evolution equations are
\begin{subequations}
\begin{eqnarray}
\frac{\ud}{\ud t}\sB_i&=&\sB_i\times(\muV{,i}\HV+\HA+\mu_\nu\Sb),
\label{eq:nfis-ve}\\
\frac{\ud}{\ud t}\Sb&=&\sum_i\muV{,i}n_{\nu,i}\sB_i\times\HV+\Sb\times\HA.
\label{eq:tnfis-ve}
\end{eqnarray}
\end{subequations}
First we assume a fixed matter background with net electron number density
 $n_e$. For high $n_e$, corresponding to $|\HA|\gg|\langle\muV{}\rangle|$, 
Eqs.~\eqref{eq:nfis-ve} and \eqref{eq:tnfis-ve} reduce to
\begin{subequations}
\begin{eqnarray}
\frac{\ud}{\ud t}\sB_i&\simeq&\sB_i\times(\HA+\mu_\nu\Sb),\\
\frac{\ud}{\ud t}\Sb&\simeq&\Sb\times\HA.
\end{eqnarray}
\end{subequations}
The above equations correspond to perfectly synchronized flavor oscillations:
in a frame rotating with an angular velocity of $-\HA$,
the total NFIS stays fixed and the individual NFIS's precess
around it with a common frequency $|\mu_\nu\Sb|$. However,
for neutrinos and antineutrinos initially in pure flavor
eigenstates, $\sB_i$ and $\Sb$ start out 
aligned or anti-aligned with 
$\HA=-\basef{z}\sqrt{2}\GF n_e$. Therefore, the above
perfect synchronized flavor oscillations reduce to a trivial case where
all $\sB_i$'s remain in the initial state (\textit{i.e.}, 
all neutrinos stay in their initial flavor states).
This trivial case is of no interest to us and will not be discussed further. 

For $|\HA|\sim|\langle\muV{}\rangle|$ and 
$|\mu_\nu\Sb|\gg|\langle\muV{}\rangle|$, the discussion is similar to the case 
with no matter background. All $\sB_i$'s precess around $\Sb$ with a 
frequency $|\mu_\nu\Sb|$ and Eq.~\eqref{eq:tnfis-ve} becomes
\begin{equation}
\frac{\ud}{\ud t}\Sb\simeq\Sb\times(\langle\muV{}\rangle\HV+\HA).
\label{eq:eom-S-matt}
\end{equation}
Therefore, the total NFIS of the gas precesses around the effective field
$\Hb^\eff=\langle\muV{}\rangle\HV+\HA$ 
and behaves just as does a single NFIS with $\sB=\Sb/(2|\Sb|)$
and $\muV{}=\langle\muV{}\rangle$
in the same matter background [see Eq.~\eqref{eq:prec}].
For the cases with $\delta m^2>0$ and $\langle\muV{}\rangle>0$ or
with $\delta m^2<0$ and $\langle\muV{}\rangle<0$,
this representative NFIS corresponds to a neutrino with energy
\begin{equation}
E_{\mathrm{sync}}\equiv\left|\frac{\delta m^2}{2\langle\muV{}\rangle}\right|.
\label{eq:Esync}
\end{equation}
For the other cases,
this representative NFIS corresponds to an antineutrino with energy 
$E_{\mathrm{sync}}$.
For an initially pure $\nu_e$ neutrino gas,
$E_{\mathrm{sync}}^{-1}$ is simply the 
neutrino energy distribution-averaged value of $E_{\nu_e}^{-1}$:
\begin{equation}
E_{\mathrm{sync}}^{-1} = \int \frac{f_{\nu_e}(E)}{E} \ud E,
\end{equation}
where $f_{\nu_e}(E)$ is the energy distribution of $\nu_e$.
For more general cases, $E_{\mathrm{sync}}$
is evaluated using Eqs.~\eqref{eq:wsync} and
\eqref{eq:Esync} with all the relevant
neutrino and antineutrino energy distributions.

The above discussion can be extended to the case of a 
slowly varying matter background in a straightforward manner.
We note that this is again an adiabatic process as discussed in
Sec.~\ref{sec:general-formalism} except that $\Sb$ takes the
place of $\sB_\nu$ this time. The angle between $\Sb$ and $\Hb^\eff$
is therefore constant. A gas of initially dominantly 
$\nu_e$ with $|\mu_\nu\Sb|\gg|\langle\muV{}\rangle|$ acts just like 
a single neutrino
with energy $E_{\mathrm{sync}}$ propagating in this matter background.
For a normal mass hierarchy ($\delta m^2>0$), there may be an MSW 
resonance that can enhance flavor transformation. In contrast, no
MSW resonance exists and flavor transformation is suppressed by the matter
effect for an inverted mass hierarchy ($\delta m^2<0$). 

Obviously, for a neutrino and/or antineutrino gas
 with $|\mu_\nu\Sb|\ll|\langle\muV{}\rangle|$, 
there is no synchronized flavor transformation.

\section{Bi-Polar Flavor Transformation%
\label{sec:bi-polar}}
The astrophysical environments where neutrino flavor
transformation is of interest do not
always provide conditions which are favorable for synchronization.
For a neutrino gas to be in the synchronized mode, the neutrinos have
to be prepared in such a way that the corresponding
NFIS's are strongly aligned in one direction. 
There are important regimes where this does not occur.

For example, consider the $2\times 2$ mixing channels
$\nu_e\rightleftharpoons\nu_\tau$ and
$\bar\nu_e\rightleftharpoons\bar\nu_\tau$ in the late-time,
shocked region above the proto-neutron star.
By definition, $\sB_i=\basef{z}/2$ for a $\nu_e$ or $\bar\nu_\tau$ and
$\sB_i=-\basef{z}/2$ for a $\bar\nu_e$ or $\nu_\tau$. Therefore
$\nu_e$, $\bar\nu_e$, $\nu_\tau$, and $\bar\nu_\tau$ 
form two NFIS blocks pointing in opposite
directions when they leave the neutrino sphere.
The subsequent behavior of these neutrinos is interesting.
We show below that under the right conditions large-scale
collective ``swapping'' of flavors 
$\nu_e\rightleftharpoons\nu_\tau$ and
$\bar\nu_e\rightleftharpoons\bar\nu_\tau$
can occur in a mode in which the NFIS blocks remain
more or less oppositely-directed. This is an example
of the bi-polar mode.

In Ref.~\cite{Kostelecky:1993dm}, numerical simulations
of a homogeneous, dense neutrino-antineutrino gas in the
absence of a matter background showed that the flavor
``swapping'' in the bi-polar mode occurred at a higher
frequency than would vacuum oscillations. 
Ref.~\cite{Kostelecky:1994dt} gave an analytical solution to
a simple bi-polar system, a gas initially consisting of equal numbers 
of mono-energetic $\nu_e$ and $\bar\nu_e$,  for 
a normal mass hierarchy.
Ref.~\cite{Samuel:1995ri} generalized the solution to a gas of 
unequal numbers of $\nu_e$ and $\bar\nu_e$ with different energies, again
for a normal mass hierarchy scenario, and found that
the system exhibits bimodal features (dual frequencies).

In this section we again  adopt a physical, analytical
 approach and use a simple example to illustrate 
neutrino flavor transformation in bi-polar systems from the
energy conservation perspective. 
For the first time,
we explain why large flavor mixing can develop in some bi-polar
systems even with a small mixing angle.
We will then extend the discussion to more general bi-polar
systems using the co-rotating frame, and discuss how bimodal
features can appear in such systems. We will propose 
criteria under which a NFIS system can be in the bi-polar mode, and show that
a bi-polar system is at least semi-stable. We will conclude
this section with some discussion on the effects of the matter
background on bi-polar flavor transformation.

\subsection{A Simple Example of Bi-Polar Flavor Transformation%
\label{sec:simple-bi-polar}}
We start with a simple bi-polar system initially consisting of mono-energetic 
$\nu_e$ and $\bar\nu_e$ with an equal number density $n_\nu$, which
form two NFIS blocks $\Sb_1(0)=n_\nu\sB_{\nu_e}=\basef{z}n_\nu/2$ and 
$\Sb_2(0)=n_\nu\sB_{\bar\nu_e}=-\basef{z}n_\nu/2$. This system is uniform and
isotropic and has no matter background. The evolution of $\Sb_1$
and $\Sb_2$ is governed by [see Eq.~\eqref{eq:nfisev}]
\begin{subequations}
\label{eq:eom-bi-polar}
\begin{eqnarray}
\frac{\ud}{\ud t}\Sb_1 &=&\Sb_1\times(\muV{,1}\HV+\mu_\nu\Sb_2), \\
\frac{\ud}{\ud t}\Sb_2 &=&\Sb_2\times(\muV{,2}\HV+\mu_\nu\Sb_1),
\end{eqnarray}
\end{subequations}
where $\muV{,1}=-\muV{,2}=\muV{}$. With the definition of
\begin{equation}
\Sb_+\equiv \Sb_1+\Sb_2 \quad \text{and} \quad \Sb_-\equiv \Sb_1-\Sb_2,
\label{eq:SpSm-def}
\end{equation}
we find
\begin{subequations}
\label{eq:SpSm-eom}
\begin{eqnarray}
\frac{\ud}{\ud t}\Sb_+ &=&\muV{}\Sb_-\times\HV, 
\label{eq:Sp-eom} \\
\frac{\ud}{\ud t}\Sb_- &=&\muV{}\Sb_+\times\HV 
+ \mu_\nu \Sb_- \times \Sb_+.
\label{eq:Sm-eom}
\end{eqnarray}
\end{subequations}
The initial conditions are
\begin{subequations}
\begin{eqnarray}
\Sb_+(0)&=&0,\\
\Sb_-(0)&=&n_\nu \basef{z}
=n_\nu(\basev{x}\sin2\thetav+\basev{z}\cos2\thetav),
\end{eqnarray}
\end{subequations}
where $\basev{x}$ and $\basev{z}$ are the unit vectors in the $x$- and 
$z$-directions, respectively, in the vacuum mass basis ($\basev{z}\equiv\HV$).
Using these conditions and Eq.~\eqref{eq:SpSm-eom}, we can show that
\begin{equation}
\Sb_+\cdot\basev{x}=\Sb_+\cdot\basev{z}=\Sb_-\cdot\basev{y}=0.
\label{eq:SpSm-zeros}
\end{equation}
In other words, $\Sb_+$ can only move parallel to $\basev{y}$ while
$\Sb_-$ is confined to move in the plane defined by $\basev{x}$ and $\basev{z}$.

The evolution of $\Sb_+$ and $\Sb_-$ obeys conservation of
the total effective energy
\begin{subequations}
\begin{align}
\mathcal{E} &= -\muV{,1} \Sb_1\cdot\HV -\muV{,2} \Sb_2\cdot\HV 
\nonumber\\
&\quad- \frac{\mu_\nu}{2}(\Sb_1+\Sb_2)^2
\label{eq:E-bi-polar}\\
&=-\muV{}\Sb_-\cdot\HV - \frac{\mu_\nu}{2}\Sb_+^2=-\muV{}n_\nu\cos2\thetav,
\end{align}
\end{subequations}
which gives
\begin{equation}
|\Sb_-|\cos\vartheta=n_\nu\cos2\thetav-\frac{\mu_\nu}{2\muV{}}\Sb_+^2
\label{eq:SpSm-conserv1}
\end{equation}
with $\vartheta$ being the angle between $\Sb_-$ and $\basev{z}$.
Further, it can be shown from Eq.~\eqref{eq:SpSm-def} that
\begin{equation}
\Sb_+^2+\Sb_-^2=n_\nu^2.
\label{eq:SpSm-conserv2}
\end{equation}
Combining Eqs.~\eqref{eq:SpSm-conserv1} and \eqref{eq:SpSm-conserv2}, 
we obtain
\begin{equation}
\cos\vartheta=\frac{\cos2\thetav}{s_-}-
\frac{\mu_\nu n_\nu}{2\muV{}}\left(\frac{1}{s_-}-s_-\right),
\label{eq:vartheta}
\end{equation}
where
\begin{equation}
s_-\equiv|\Sb_-|/n_\nu.
\end{equation}
Noting that $\mu_\nu=-2\sqrt{2}\GF<0$ and $s_-\leq 1$, 
we see that for a normal mass
hierarchy ($\muV{}>0$), $\cos2\thetav\leq\cos\vartheta\leq 1$ and
$\Sb_-$ is constrained to oscillate around
$\basev{z}$ with $-2\thetav\leq\vartheta\leq 2\thetav$. 
For $\thetav\ll 1$, the system stays close
to the initial state and there is very little flavor mixing.

The situation for an inverted mass hierarchy ($\muV{}<0$) is more 
complicated. We proceed by first rewriting Eq.~\eqref{eq:vartheta} as
\begin{subequations}
\label{eq:vartheta2}
\begin{eqnarray}
\cos\vartheta&=&\left(\cos2\thetav-\frac{\mu_\nu n_\nu}{2\muV{}}\right)
\frac{1}{s_-}+\left(\frac{\mu_\nu n_\nu}{2\muV{}}\right)s_-\\
&=&\cos2\thetav\left[\left(1-\frac{n_\nu}{n_\nu^{\rm cri}}\right)
\frac{1}{s_-}+\left(\frac{n_\nu}{n_\nu^{\rm cri}}\right)s_-\right],
\end{eqnarray}
\end{subequations}
where
\begin{equation}
n_\nu^{\rm cri}\equiv\frac{2\muV{}}{\mu_\nu}\cos2\thetav
\end{equation}
is a positive characteristic neutrino number density for $\muV{}<0$.
The evolution of the bi-polar system under consideration falls into the
following three categories depending on the parameter 
$n_\nu/n_\nu^{\rm cri}$.
For $n_\nu/n_\nu^{\rm cri}\leq 1/2$, $\ud\cos\vartheta/\ud s_-\leq 0$
and $\Sb_-$ is constrained to oscillate around $\basev{z}$ with 
$-2\thetav\leq\vartheta\leq 2\thetav$ just as in the case of a normal 
mass hierarchy. In this case, the difference between
$\muV{,1}$ and $\muV{,2}$ is too large for the two NFIS blocks
to maintain strong correlation, and neutrinos and antineutrinos
oscillate as two separate sectors.
For $1/2<n_\nu/n_\nu^{\rm cri}<1$, 
$\ud\cos\vartheta/\ud s_-$ is positive for $s_-\sim 1$ but becomes 0
and then negative for smaller $s_-$, and the maximum value
$\vartheta_{\rm max}$ corresponding to $\ud\cos\vartheta/\ud s_-=0$
is given by
\begin{equation}
\cos\vartheta_{\rm max}=2\cos2\thetav\sqrt{\frac{n_\nu}{n_\nu^{\rm cri}}
\left(1-\frac{n_\nu}{n_\nu^{\rm cri}}\right)}.
\end{equation}
In this case, $\vartheta$ first increases from $2\thetav$ for the initial 
state to $\vartheta_{\rm max}$ as $s_-$ decreases and then decreases 
to 0 as $s_-$ further decreases to its minimum value. Subsequently the 
motion of $\Sb_-$ is mirrored in the other half of the plane defined by 
$\basev{x}$ and $\basev{z}$. 
Note that for each complete cycle $\Sb_-$ 
reaches the position at $\vartheta=2\thetav$ ($-2\thetav$) twice but with 
$s_-=1$ and a smaller value, respectively. 
For $n_\nu/n_\nu^{\rm cri}\geq 1$, Eq.~\eqref{eq:vartheta2} shows that 
$\ud\cos\vartheta/\ud s_-$ is always positive,
$-1\leq\cos\vartheta\leq\cos2\thetav$ and $\Sb_-$ oscillates around 
$\basev{z}$ with $2\thetav\leq\vartheta\leq 2\pi-2\thetav$ (note that
for $n_\nu/n_\nu^{\rm cri}=1$, $\Sb_-$ shrinks to 0 at $\vartheta=\pi/2$
and $3\pi/2$, therefore appearing to skip the range $\pi/2<\vartheta<3\pi/2$).  
In the limit $n_\nu/n_\nu^{\rm cri}\gg 1$, $s_-$ stays $\sim 1$ as $\Sb_-$ 
rotates in the plane defined by $\basev{x}$ and $\basev{z}$. This is
particularly interesting because the two NFIS blocks remain anti-aligned 
and can completely reverse their initial directions, which means that full 
conversion of the initial $\nu_e$ and $\bar\nu_e$ occurs even for 
$\thetav\ll 1$.

\begin{figure}
\begin{center}
\includegraphics*[scale=0.8, keepaspectratio]{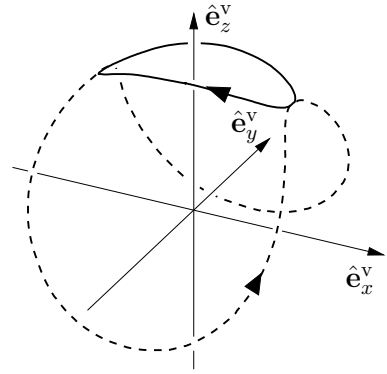}
\end{center}
\caption{\label{fig:bi-polar-curves}%
The solution of $\sB_1$ for a simple bi-polar system
in the vacuum mass basis for
a normal (solid line) and an inverted (dashed line) mass
hierarchy, respectively. The solution of $\sB_2$ can be obtained
from that of
$\sB_1$ using Eq.~(\ref{eq:SpSm-nonzeros}).}
\end{figure}

\begin{figure*}
\begin{center}
$\begin{array}{@{}c@{\hspace{\myfigsep}}c@{}}
\includegraphics*[width=\myfigwid, keepaspectratio]{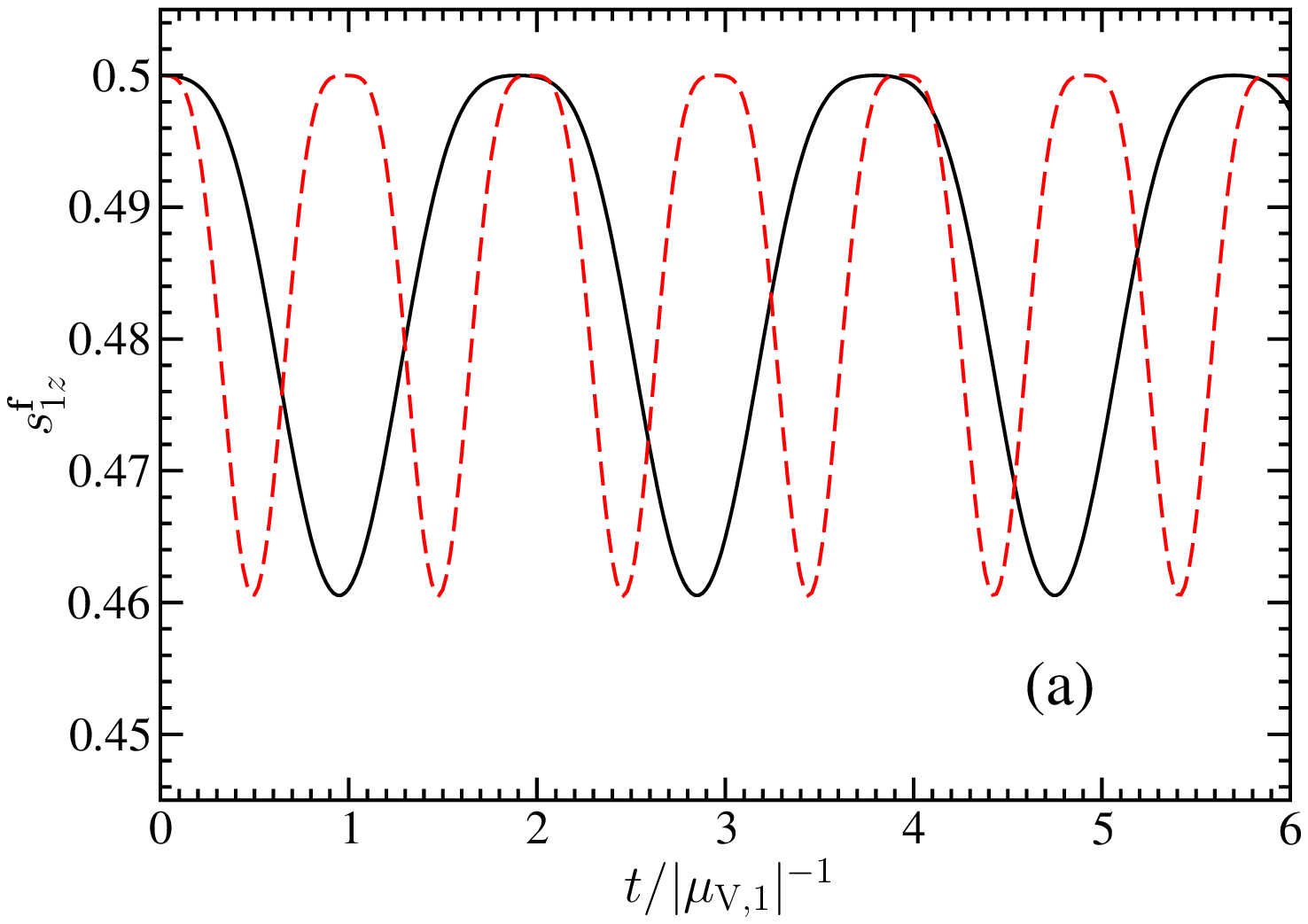} &
\includegraphics*[width=\myfigwid, keepaspectratio]{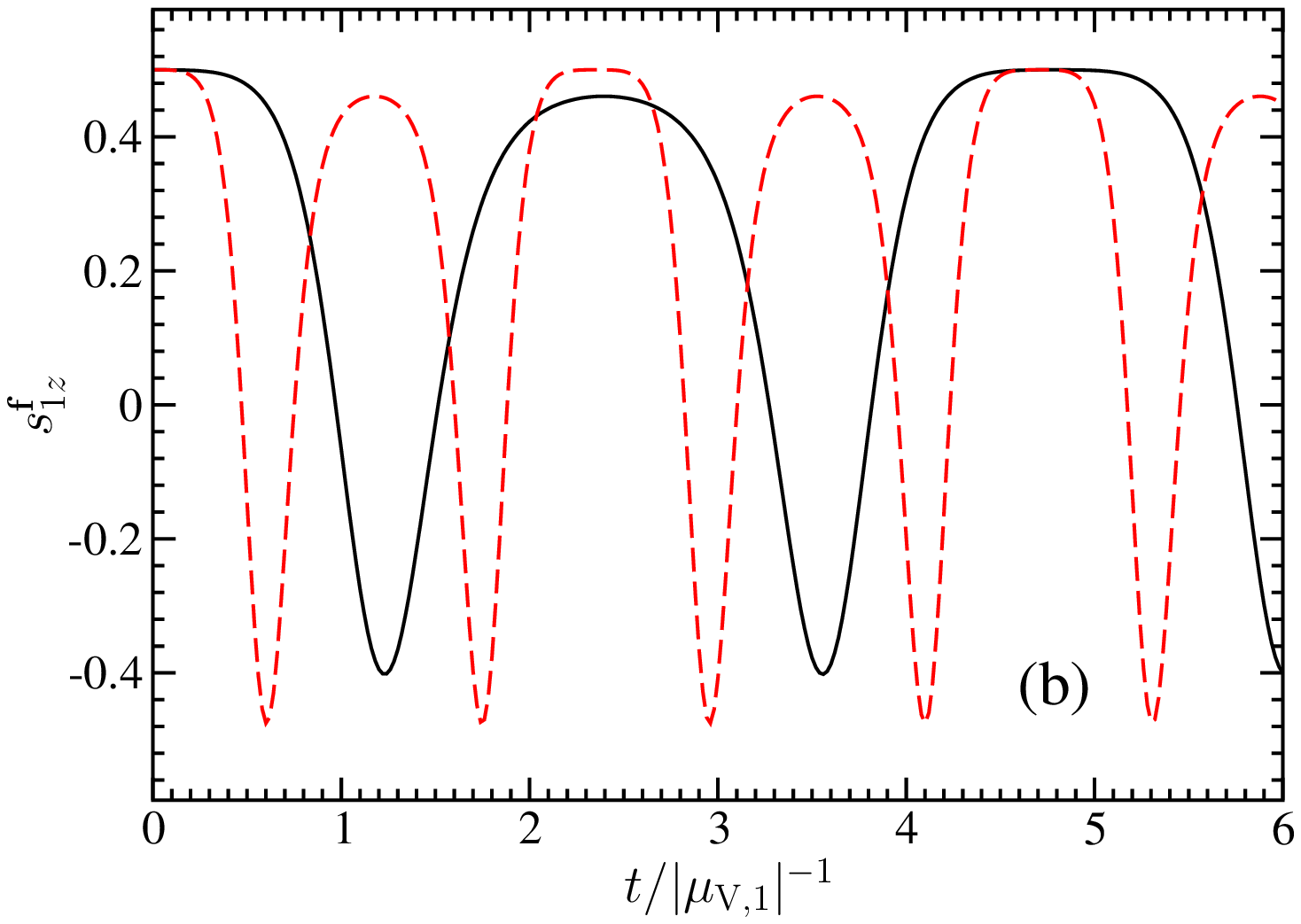}
\end{array}$
\end{center} 
\caption{\label{fig:simple-bi-polar}(Color online)
The evolution of the $z$-component of $\sB_1$ 
in the flavor basis for the simple
bi-polar system with (a) a normal and (b) an inverted mass hierarchy.
The solid lines are for $|\muV{,1}|/|\mu_\nu n_\nu|=1/10$, and
the dashed lines are for $|\muV{,1}|/|\mu_\nu n_\nu|=1/40$. The vacuum
mixing angle $\thetav$ is taken to be 0.1.}
\end{figure*}

The above results on the evolution of $\Sb_-$ and $\Sb_+$
can be adapted easily to describe the evolution of the 
individual NFIS's $\sB_1$ and $\sB_2$ of an initial $\nu_e$ and
$\bar\nu_e$, respectively.
Using Eqs.~\eqref{eq:SpSm-def} and \eqref{eq:SpSm-zeros},
we find 
\begin{subequations}
\label{eq:SpSm-nonzeros}
\begin{eqnarray}
s_{1x}^\mathrm{v} = -s_{2x}^\mathrm{v} &=&\frac{\Sb_-\cdot\basev{x}}{2n_\nu}
=\frac{s_-\sin\vartheta}{2},\\ 
s_{1y}^\mathrm{v} = s_{2y}^\mathrm{v} &=&\frac{\Sb_+\cdot\basev{y}}{2n_\nu},\\ 
s_{1z}^\mathrm{v} = -s_{2z}^\mathrm{v} &=&\frac{\Sb_-\cdot\basev{z}}{2n_\nu}
=\frac{s_-\cos\vartheta}{2},
\end{eqnarray}
\end{subequations}
where for example, $s_{1x}^\mathrm{v}$ is the $x$-component of $\sB_1$
in the vacuum mass basis. As a visual illustration, we show in
Fig.~\ref{fig:bi-polar-curves} the evolution of $\sB_1$ in the vacuum mass
basis
for the case of a normal mass hierarchy (solid curve)
and for the case of an inverted mass hierarchy with $n_\nu/n_\nu^{\rm cri}>1$
(dashed curve). The trajectory of $\sB_1$ for each 
case marks the intersection between the parabolic surface representing
\begin{equation}
\muV{}\left(s_{1z}^\mathrm{v}-\frac{\cos2\thetav}{2}\right)
+\mu_\nu n_\nu (s_{1y}^\mathrm{v})^2=0,
\label{eq:parabola}
\end{equation}
which is equivalent to Eq.~\eqref{eq:SpSm-conserv1}, and the spherical surface
representing
\begin{equation}
(s_{1x}^\mathrm{v})^2 + (s_{1y}^\mathrm{v})^2 + (s_{1z}^\mathrm{v})^2
= \left(\frac{1}{2}\right)^2,
\end{equation}
which follows from the fixed magnitude of $\sB_1^2=(1/2)^2$.
The evolution of $\sB_2$ can be obtained from that of $\sB_1$
based on Eq.~\eqref{eq:SpSm-nonzeros}.

Of course, the flavor evolution of an initial $\nu_e$ is described most directly
by the $z$-component $s_{1z}^\mathrm{f}$ of $\sB_1$ in the flavor basis.
The unit vectors in this basis are related to those in the vacuum mass
basis as
\begin{subequations}
\begin{eqnarray}
\basef{x}&=&\basev{x}\cos2\thetav-\basev{z}\sin2\thetav,\\
\basef{y}&=&\basev{y},\\
\basef{z}&=&\basev{x}\sin2\thetav+\basev{z}\cos2\thetav.
\end{eqnarray}
\end{subequations}
From the above equations and Eq.~\eqref{eq:SpSm-nonzeros}, we obtain
\begin{equation}
s_{1z}^\mathrm{f}=s_{1x}^\mathrm{v}\sin2\thetav+
s_{1z}^\mathrm{v}\cos2\thetav=\left(\frac{s_-}{2}\right)\cos(\vartheta-2\thetav).
\label{eq:s1zf}
\end{equation}
As $\Sb_-$ evolves very little from the initial state for $\thetav\ll 1$ in
the case of a normal mass hierarchy and in the case
of an inverted mass hierarchy with $n_\nu/n_\nu^{\rm  cri}\leq 1/2$,
there is little flavor evolution in these cases. In contrast, 
for the case of an 
inverted mass hierarchy with $n_\nu/n_\nu^{\rm  cri}\gg1$, we have
$s_-\sim 1$ and $2\thetav\leq\vartheta\leq 2\pi-2\thetav$, 
so an initial $\nu_e$ 
can be converted essentially fully into a $\nu_\tau$ even for $\thetav\ll 1$. 

Taking $\thetav=0.1$ and 
$|\mu_\nu| n_\nu/|\muV{}|=10$ ($n_\nu/n_\nu^{\rm cri}=5.1$ for 
$\muV{}<0$), we show the time evolution of $s_{1z}^\mathrm{f}$ as the 
solid lines in Figs.~\ref{fig:simple-bi-polar}(a) and 
\ref{fig:simple-bi-polar}(b)
for a normal and an inverted mass hierarchy, respectively. The cases with
the same $\thetav$ but $|\mu_\nu| n_\nu/|\muV{}|=40$ 
($n_\nu/n_\nu^{\rm cri}=20.4$ for $\muV{}<0$) are shown as the dashed 
lines. [In order to show the small evolution in the case of a normal
mass hierarchy, we have greatly expanded the vertical scale in 
Fig.~\ref{fig:simple-bi-polar}(a).]

We note that the period of vacuum oscillations in these numerical
examples is $T_\mathrm{vac}=2\pi/|\muV{}|$.
This is longer than 
the bi-polar oscillation periods $T_\mathrm{bi}$ 
shown in Fig.~\ref{fig:simple-bi-polar}.
In addition, the bi-polar oscillation periods decrease by a factor
of 2 when the neutrino density $n_\nu$ is increased by a factor of 4.
These observations can be understood from Eq.~\eqref{eq:SpSm-eom},
even without an outright solution of this equation.
In the limit $|\mu_\nu| n_\nu/|\muV{}|\gg 1$,
the second term on the right hand side of Eq.~\eqref{eq:Sm-eom}
dominates, and $\Sb_-$ simply rotates around $\Sb_+$ 
with roughly a constant magnitude $n_\nu$ and frequency
\begin{equation}
T_\mathrm{bi}^{-1} \sim |\mu_\nu| \langle |\Sb_+|\rangle.
\label{eq:Tbi-1}
\end{equation}
The average value of $|\Sb_+|$ in the above equation can be
estimated from Eq.~\eqref{eq:Sp-eom}:
\begin{equation}
\frac{\langle |\Sb_+|\rangle }{T_\mathrm{bi}}
\sim |\muV{}\Sb_-|\simeq |\muV{}|n_\nu.
\label{eq:Sm-avg}
\end{equation}
Combining Eqs.~\eqref{eq:Tbi-1} and \eqref{eq:Sm-avg} we obtain
\begin{equation}
T_\mathrm{bi}\sim \frac{1}{\sqrt{|\muV{}\mu_\nu|n_\nu}}.
\label{eq:Tbi}
\end{equation}
This simple dimensional analysis agrees with the exact expression for the
bi-polar period in the normal mass hierarchy case
\cite{Kostelecky:1994dt}.
Therefore, for a large neutrino density $n_\nu\gg |\muV{}/\mu_\nu|$,
the bi-polar oscillation period
$T_\mathrm{bi}$ is much smaller than $T_\mathrm{vac}$ and
decreases as $\sim 1/\sqrt{n_\nu}$.

\subsection{General Bi-Polar Systems%
\label{sec:general-bi-polar}}

We next look at a slightly more complicated neutrino-antineutrino system.
In particular, we consider a system which 
is the same as that discussed in Sec.~\ref{sec:simple-bi-polar}
except that $\nu_e$ and $\bar\nu_e$ have different energies.
We again define $\Sb_+$ and $\Sb_-$ as in Eq.~\eqref{eq:SpSm-def}.
Using Eq.~\eqref{eq:eom-bi-polar}, we find
\begin{subequations}
\label{eq:SpSm-eom-gen}
\begin{eqnarray}
\frac{\ud}{\ud t}\Sb_+ &=& \Sb_+ \times \Hb_+ + \Sb_-\times\Hb_-, 
\label{eq:Sp-eom-gen} \\
\frac{\ud}{\ud t}\Sb_- &=& \Sb_- \times \Hb_+ + \Sb_+\times\Hb_- 
+ \mu_\nu \Sb_- \times \Sb_+,
\label{eq:Sm-eom-gen}
\end{eqnarray}
\end{subequations}
where 
\begin{equation}
\Hb_\pm \equiv\frac{\muV{,1} \pm \muV{,2}}{2} \HV.
\end{equation}
When viewed in the reference frame rotating with
angular velocity $-\Hb_+$, 
the e.o.m.~of $\widetilde{\Sb}_+$ and $\widetilde{\Sb}_-$
derived from Eq.~\eqref{eq:SpSm-eom-gen} are exactly the same
as that in Eq.~\eqref{eq:SpSm-eom} and must, therefore, have
the same solution. When viewed in the lab frame, this NFIS system
not only demonstrates the bi-polar oscillation as discussed in 
Sec.~\ref{sec:simple-bi-polar}, but also  rotates around $\HV$
at the same time. We regard this kind of flavor transformation
as also being of bi-polar type. Note that this bi-polar system
has two intrinsic periods, \textit{i.e.}, $T_\mathrm{bi}$ and $2\pi/|\Hb_+|$.
This bimodal feature of the neutrino-antineutrino system
was first discussed in Ref.~\cite{Samuel:1995ri}.

\begin{table}
\caption{\label{tab:bi-polar-large-mixing}The conditions
for a dense neutrino gas starting as two groups of mono-energetic 
neutrino species with equal number to develop large flavor mixing in
the small mixing angle scenario. Combinations of
neutrino species other than those shown here
will not develop large flavor mixing in this case.}
\begin{ruledtabular}
\begin{tabular}{c|cccc}
&$\nu_e$--$\bar\nu_e$ & $\bar\nu_\tau$--$\nu_\tau$
& $\nu_e$--$\nu_\tau$ & $\bar\nu_\tau$--$\bar\nu_e$\\
\hline
$\delta m^2 > 0$ & Never & Always 
& $E_{\nu_e} > E_{\nu_\tau}$ & $E_{\bar\nu_\tau} < E_{\bar\nu_e}$   \\
$\delta m^2 < 0$ & Always & Never
& $E_{\nu_e} < E_{\nu_\tau}$ & $E_{\bar\nu_\tau} > E_{\bar\nu_e}$ 
\end{tabular}
\end{ruledtabular}
\end{table}

Note that the above arguments employing co-rotating frames 
apply not only to systems
consisting of $\nu_e$--$\bar\nu_e$, but also to systems of
$\bar\nu_\tau$--$\nu_\tau$, $\nu_e$--$\nu_\tau$ or
$\bar\nu_\tau$--$\bar\nu_e$. Because $\nu_e$--$\bar\nu_e$ systems
can develop large flavor mixing in the case of a small mixing angle
and an inverted mass hierarchy, the other systems also can exhibit 
 the same phenomenon as long as 
\begin{equation}
\muV{,1} < \muV{,2},
\end{equation}
where $\muV{,1}$ and $\muV{,2}$ are the vacuum coupling coefficients
of the ``spin-up'' and ``spin-down'' NFIS's, respectively. 
For convenience, we have listed these
conditions in Table~\ref{tab:bi-polar-large-mixing}.

\begin{figure*}
\begin{center}
$\begin{array}{@{}c@{\hspace{\myfigsep}}c@{}}
\includegraphics*[width=\myfigwid, keepaspectratio]{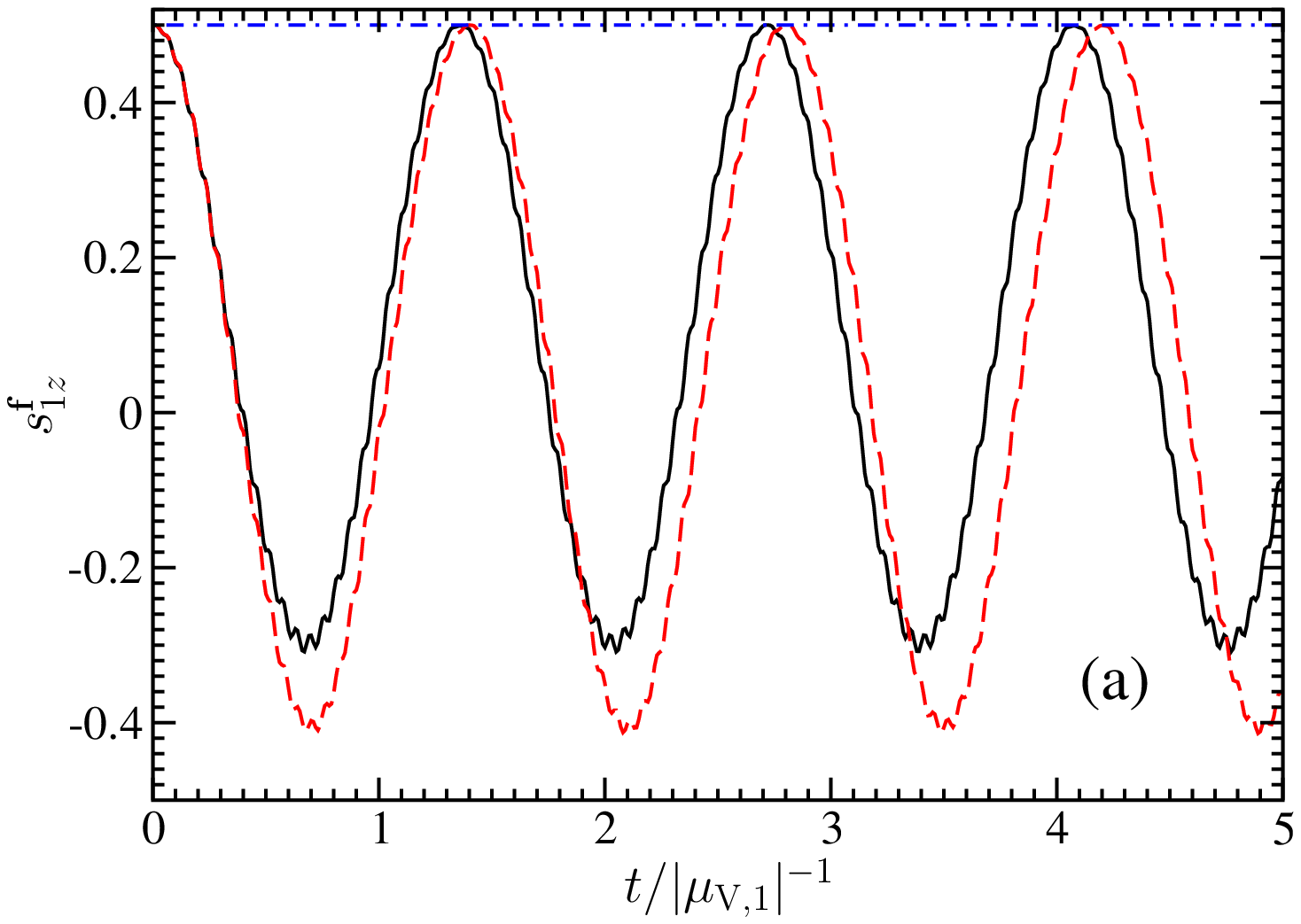} &
\includegraphics*[width=\myfigwid, keepaspectratio]{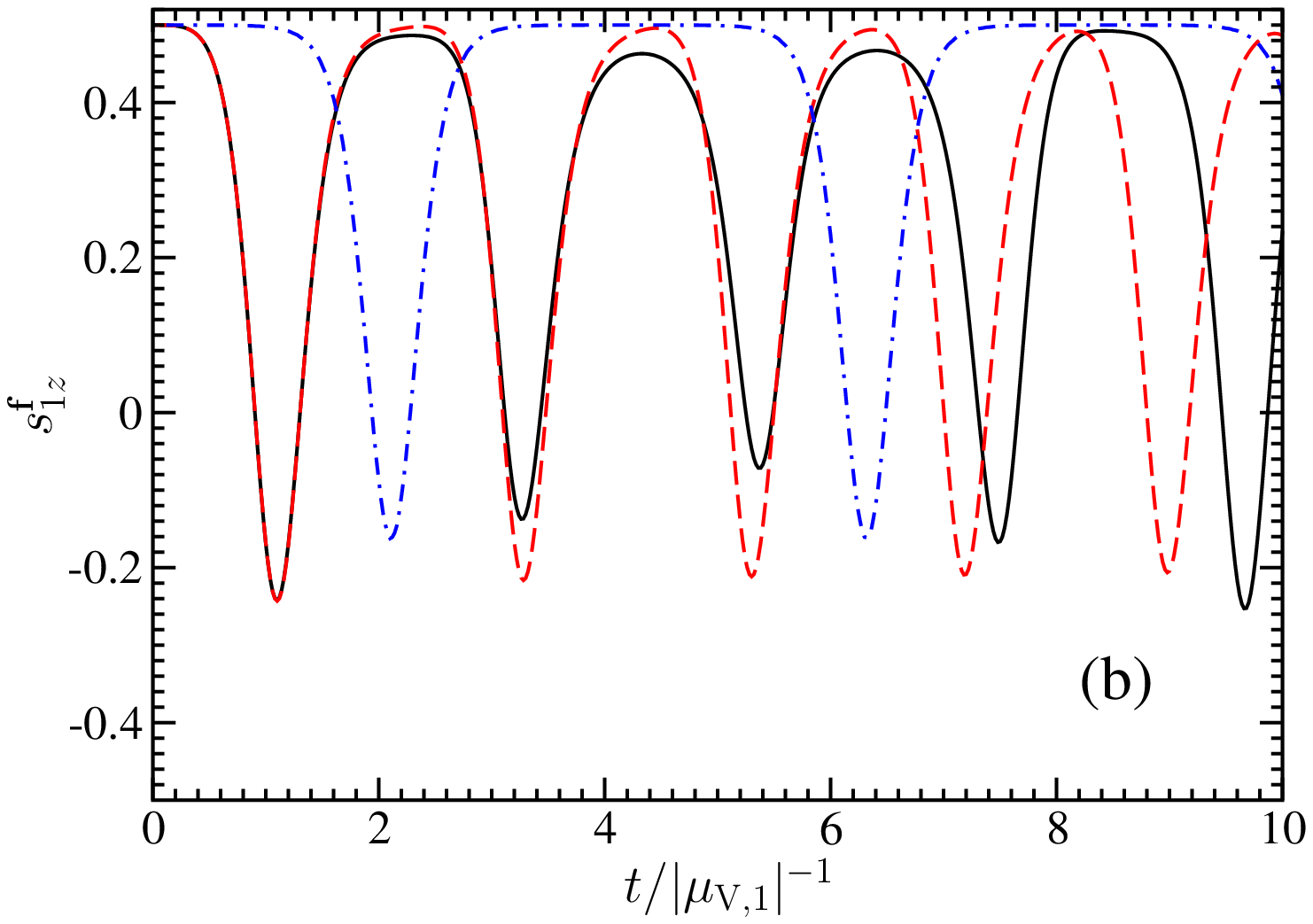}
\end{array}$
\end{center}
\caption{\label{fig:bi-polar-eg}(Color online)
The evolution of the $z$-component of $\sB_1$ 
(in the flavor basis) for representative bi-polar systems.
These systems consist of
mono-energetic neutrinos in initially pure $\nu_e$ and $\bar\nu_e$ flavor
states.
The mixing parameters $\thetav$ and $\delta m^2$ are
$0.56$ and $8\times 10^{-5}\,\mathrm{eV}^2$, respectively,
for the calculations
in panel (a), and are $0.1$ and $-3\times 10^{-3}\,\mathrm{eV}^2$,
respectively,
for panel (b). The energies of $\nu_e$ and $\bar\nu_e$
are taken to be $11\,\mathrm{MeV}$ and $16\,\mathrm{MeV}$, respectively. 
The (effective) number
densities of $\nu_e$ and $\bar\nu_e$ are $10^{28}\,\mathrm{cm}^{-3}$
and $6.9\times 10^{27}\,\mathrm{cm}^{-3}$, respectively.
The solid, dashed, and dot-dashed lines are plotted with electron
number density taken to be $0$, $10^{26}$, and $10^{29}\,\mathrm{cm}^{-3}$,
respectively.}
\end{figure*}

From the simple examples discussed above we can infer a general
description of a system possessing bi-polar oscillations:
a system  composed of two groups of neutrinos and/or antineutrinos
of roughly equal numbers, where
 the corresponding NFIS's point in two roughly opposite directions
and have different characteristic
values of $\muV{}$.

In Fig.~\ref{fig:bi-polar-eg} (the solid lines) we illustrate two
non-ideal bi-polar systems. These examples consist of gases of initially pure
$\nu_e$ and $\bar\nu_e$  with $n_{\nu_e}\neq n_{\bar\nu_e}$
and $E_{\nu_e}\neq E_{\bar\nu_e}$. One of the examples
[Fig.~\ref{fig:bi-polar-eg}(a)] illustrates neutrino mixing
with a large mixing angle and $\delta m^2\simeq \delta m_\odot^2$,
and the other [Fig.~\ref{fig:bi-polar-eg}(b)]
uses a small mixing angle and $\delta m^2\simeq -\delta m_\mathrm{atm}^2$.

As the difference between the
densities of the two NFIS blocks becomes larger and larger,
one of the NFIS blocks will eventually dominate the other, and
the system will become synchronized rather than bi-polar.
This can been seen from Eq.~\eqref{eq:SpSm-eom-gen}.
For simplicity, we work in the  frame rotating with angular velocity
$-\Hb_+$ where the e.o.m.~of $\widetilde{\Sb}_\pm$
take the same form as Eq.~\eqref{eq:SpSm-eom}. 
We note that a key characteristic of any
bi-polar system is a configuration in which
a large and near constant magnitude  $\widetilde{\Sb}_-$ vector 
rotates about $\widetilde{\Sb}_+$ [Eq.~\eqref{eq:Sm-eom}]. 
In this configuration, $\widetilde{\Sb}_+$ typically has a small,
variable magnitude.
If one of the NFIS blocks dominates, 
it is possible that $|\mu_\nu\widetilde{\Sb}_+|$
 will be bigger than 
$|\dot{\widetilde{\Sb}}_+\times\widetilde{\Sb}_+|/|\widetilde{\Sb}_+|^2$,
and the adiabatic condition can be satisfied.
(Note that the rate of change of $\widetilde{\Sb}_+$
is bounded by the intrinsic frequency of the bi-polar oscillation.)
If the adiabatic condition is satisfied, 
then $\widetilde{\Sb}_-$ will  precess rapidly
around $\widetilde{\Sb}_+$, with a constant relative angle between them.
In this case,  $\widetilde{\Sb}_-$ will
average out to be 
$(\widetilde{\Sb}_-\cdot\widetilde{\Sb}_+)\widetilde{\Sb}_+
/|\widetilde{\Sb}_+|^2$.
At the same time,
$\widetilde{\Sb}_+$ will have roughly constant magnitude
and will rotate around $\HV$ with a angular frequency
 [Eq.~\eqref{eq:Sp-eom}]
\begin{equation}
\tilde\omega_+\equiv \left(
\frac{\muV{,1}-\muV{,2}}{2}\right)
\frac{\widetilde{\Sb}_-\cdot\widetilde{\Sb}_+}{|\widetilde{\Sb}_+|^2}.
\end{equation}
Clearly, the dominant oscillation behavior of the NFIS system
is the slow rotation around $\HV$ and the synchronized mode obtains
in this case. Indeed, one can explicitly show that $\tilde\omega_+$ is
the synchronization frequency $\tilde\omega_{\mathrm{sync}}$
of the system in the co-rotating frame. 

From the above arguments one can see that the condition for a bi-polar
system to degrade into a synchronized mode is the same as that
for $\widetilde{\Sb}_-$ to adiabatically precess around $\widetilde{\Sb}_+$.
Therefore, the criterion for a NFIS system to be bi-polar is
that
\begin{equation}
|\mu_\nu \widetilde{\Sb}_+| \lesssim 
\frac{|\dot{\widetilde{\Sb}}_+\times\widetilde{\Sb}_+|}{|\widetilde{\Sb}_+|^2}
=|\tilde\omega_\mathrm{sync}|\cdot|\sin\xi|,
\end{equation}
where $\xi$ is the angle between the directions of $\widetilde{\Sb}_+$
and $\HV$. We note that this criterion is exactly the opposite
of the synchronization criterion given in Eq.~\eqref{eq:necessary-sync-cond}.
In many cases, the criterion for bi-polar flavor transformation
can be expressed as 
\begin{equation}
|\mu_\nu\Sb|\lesssim |\langle\muV{}\rangle|.
\label{eq:bi-polar-cond}
\end{equation}

We now comment on the stability of the bi-polar mode. 
Realistic systems of interest usually consist
of neutrinos and/or antineutrinos with continuous energy
distributions. Because neutrinos (antineutrinos) of different
energies have different vacuum oscillation frequencies, one might 
suspect that the bi-polar mode eventually collapses
as a result of destructive interference.
We have argued above (Sec.~\ref{sec:general-synchronization})
that the conservation of total effective energy $\mathcal{E}$
essentially guarantees the stability of the synchronized mode.
This conclusion does not extend directly to the bi-polar mode.
However, this energy conservation condition \textit{does}
provide some shielding of the
two-oppositely-directed-NFIS-block configuration 
against rapid destructive interference-driven breakdown.
This is because 
the effective energies of the two NFIS blocks
 ($-\mu_\nu\Sb_1^2/2\simeq -\mu_\nu n_{\nu,1}^2/8$
and $-\mu_\nu\Sb_2^2/2\simeq -\mu_\nu n_{\nu,2}^2/8$)
and the interaction energy of these blocks
($-\mu_\nu \Sb_1\cdot\Sb_2\simeq \mu_\nu n_{\nu,1}n_{\nu,2}/4$)
sum to almost zero, and because each of these
ingredient energies are large in magnitude and not
easily altered.
If the bi-polar configuration is ever to  break down,
the two NFIS
blocks must disassemble simultaneously in a symmetric way
in order to conserve the total effective energy.
As a result,
the bi-polar mode is at least semi-stable.
It has been observed 
that the two-oppositely-directed-NFIS-block configuration is
roughly maintained in example numerical simulations
 \cite{Kostelecky:1993yt,Kostelecky:1993ys}.

\subsection{Bi-Polar Flavor Transformation with a Matter Background%
\label{sec:bi-polar-matter}}

In Fig.~\ref{fig:bi-polar-eg} we present results of numerical
solutions to the e.o.m.~for two NFIS blocks [Eq.~\eqref{eq:nfisev}].
In this figure, we show examples
(the dashed and dot-dashed lines)
of flavor oscillations in
gases of initially pure $\nu_e$ and $\bar\nu_e$
 in the presence of various matter backgrounds.
It can be seen that flavor transformation
is suppressed for the scenario with a large mixing angle and a
normal mass hierarchy if $n_e \gtrsim n_\nu$. On the other hand,
large flavor mixing still occurs in  these systems
with small mixing angles and an inverted mass hierarchy
even if the electron density dominates, although the 
 flavor oscillation period is somewhat longer
than in the case with no matter background. 

\begin{figure}
\begin{center}
\includegraphics*[scale=0.3, keepaspectratio]{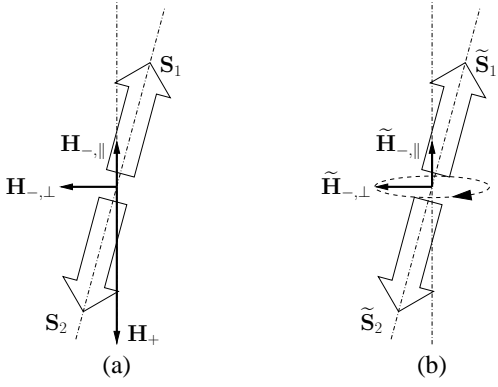}
\end{center}
\caption{\label{fig:bi-polar-matt}%
A bi-polar system in the presence of a large electron background.
(a) $\Hb_-$ can be decomposed into two components, 
$\Hb_{-,\perp}$ and
$\Hb_{-,\parallel}$, which are perpendicular and parallel to
$\Hb_+$, respectively. (b) Component $\widetilde{\Hb}_{-,\perp}$ 
will rotate very rapidly and  will have  little net
effect in the frame  rotating 
with angular velocity $-\Hb_+$.
 In the same co-rotating frame, $\widetilde{\Hb}_{-,\parallel}$
is static. This component drives the bi-polar motion of the  system.}
\end{figure}

This phenomenon can be understood
qualitatively by using the concept of co-rotating frames.
In the presence of a  matter background, the motions
of $\Sb_\pm$ still are governed by Eq.~\eqref{eq:SpSm-eom-gen}
except that $\Hb_+$ is in this case defined as
\begin{equation}
\Hb_+ = \frac{\muV{,1} + \muV{,2}}{2} \HV + \HA.
\label{eq:Hplus-matt}
\end{equation}
We decompose $\Hb_-$ into two components: $\Hb_{-,\perp}$
and $\Hb_{-,\parallel}$. These vectors are perpendicular and parallel to
$\Hb_+$, respectively.
In the reference frame rotating with angular velocity $-\Hb_+$, we have
\begin{subequations}
\begin{eqnarray}
\dot{\widetilde{\Hb}}_{-,\perp} &=& -
\widetilde{\Hb}_{-,\perp}\times\Hb_+,
\label{eq:Hminus-eom-matt}\\
\dot{\widetilde{\Hb}}_{-,\parallel} &=& 0.
\end{eqnarray}
\end{subequations}
This configuration is illustrated in Fig.~\ref{fig:bi-polar-matt}.
If $n_e$ is very large,
$\widetilde{\Hb}_{-,\perp}$ will rotate very rapidly
and the NFIS's are not able
to follow it. In this limit, $\widetilde{\Hb}_{-,\perp}$
will have on average negligible influence  on the
overall evolution of the system, at least so long as $\Sb_1$
and $\Sb_2$ are not aligned with $\widetilde{\Hb}_{-,\parallel}$.
Note that this scenario is similar to the simple
small mixing angle bi-polar example
discussed in Sec.~\ref{sec:simple-bi-polar}. However, one difference is
 that here $\widetilde{\Hb}_{-,\parallel}$ takes the place of $\muV{}\HV$
in the simple case.
Bi-polar systems with matter backgrounds, therefore, 
behave similarly to those without.

Consider the simple $\nu_e$--$\bar\nu_e$ system discussed
in  Sec.~\ref{sec:simple-bi-polar} but now with a large matter background.
The two NFIS blocks formed by neutrinos and antineutrinos are initially
aligned or anti-aligned with $\HA=\Hb_+$. 
The field component $\widetilde{\Hb}_{-,\perp}$ 
will perturb the system to break this alignment. 
The perturbation by $\widetilde{\Hb}_{-,\perp}$
is not enough to cause $\widetilde{\Sb}_1$ and $\widetilde{\Sb}_2$ to
deviate much from their original directions. This
is because $\widetilde{\Hb}_{-,\perp}$ has negligible net effects
once the angle between $\widetilde{\Sb}_1$ ($\widetilde{\Sb}_2$)
and $\widetilde{\Hb}_{-,\parallel}$ is significant. However, the configuration
with a misalignment of
$\widetilde{\Sb}_1$ and $\widetilde{\Sb}_2$ 
relative to $\widetilde{\Hb}_{-,\parallel}$
is just like the initial configuration of the simple $\nu_e$--$\bar\nu_e$
system discussed in Sec.~\ref{sec:simple-bi-polar}.
Therefore, 
for the normal mass hierarchy ($\delta m^2>0$), 
$\widetilde{\Sb}_1$ and $\widetilde{\Sb}_2$ will
oscillate around 
$\basef{z}=\HA/|\HA|=\widetilde{\Hb}_{-,\parallel}/|\widetilde{\Hb}_{-,\parallel}|$, and no significant flavor mixing occurs.
For the inverted mass hierarchy ($\delta m^2<0$),
 $\widetilde{\Sb}_1$ and $\widetilde{\Sb}_2$ could completely
swap their directions and large flavor mixing will result.

Dynamically, the configuration with $\widetilde{\Sb}_-$ aligned
with $\widetilde{\Hb}_{-,\parallel}$ is ``stable'' because 
$\widetilde{\Hb}_{-,\parallel}$
acts as a restoring ``force''  which prevents $\widetilde{\Sb}_-$  from
becoming significantly misaligned with it.
This is why neutrino-antineutrino gases of initially pure $\nu_e$
and $\bar\nu_e$ do not have large flavor mixing if $\delta m^2>0$ and 
$\thetav\ll 1$.
On the other hand, the configuration with $\widetilde{\Sb}_-$  anti-aligned
with $\widetilde{\Hb}_{-,\parallel}$ is ``unstable'' because 
$\widetilde{\Hb}_{-,\parallel}$
acts as a ``force''  which drives $\widetilde{\Sb}_-$ 
toward the (``stable'') position of alignment with 
$\widetilde{\Hb}_{-,\parallel}$ and then drives it back
to the original (``unstable'') position.
This is why neutrino-antineutrino gases of initially pure $\nu_e$
and $\bar\nu_e$ can have large flavor mixing if $\delta m^2<0$ and 
$\thetav\ll 1$.

\section{Collective Neutrino Flavor Transformation in Supernovae%
\label{sec:supernovae}}

To treat neutrino and antineutrino 
flavor transformation in a core-collapse supernova
event, we must
account for nonuniformity and anisotropy in the neutrino density 
distribution. At a radius
$r$ which is larger than  the neutrino sphere radius $R_\nu$, the neutrino
number density distribution is
\begin{equation}
\frac{\ud^2n_\nu}{\ud E_\nu\ud\Omega_\nu}=\left\{\begin{array}{ll}
\displaystyle\frac{L_\nu f_\nu(E_\nu)}{4\pi^2R_\nu^2\langle E_\nu\rangle},
&0\leq\Theta\leq\Theta_0,\\
0&\mathrm{otherwise},\end{array}\right.
\label{eq-nudens}
\end{equation}
where $L_\nu$ is the neutrino (energy) luminosity, 
$f_\nu(E_\nu)$ is the normalized neutrino energy distribution,
$\langle E_\nu\rangle$
is the average neutrino energy, $\ud\Omega_\nu$ is the differential solid angle
around the radial direction with $\Theta$ being the polar angle, and
\begin{equation}
\cos\Theta_0=\sqrt{1-(R_\nu/r)^2}.
\end{equation}

In general, flavor evolution of neutrinos traveling in different directions
above the neutrino sphere will be different due to the anisotropy of the 
neutrino density distribution. For a qualitative discussion, we will
assume the ``single-angle approximation'' 
(see, \textit{e.g.}, Ref.~\cite{Fuller:2005ae}) 
that the flavor evolution history of a radially propagating neutrino is 
representative of all neutrinos. Under this approximation,
\begin{equation}
\mu_{ij}n_{\nu,i}\to\mu_\nu n_\nu^\eff f_\nu(E_\nu)\ud E_\nu,
\end{equation}
where
\begin{subequations}
\begin{align}
n_\nu^\eff&\equiv \int(1-\cos\Theta)
\frac{\ud^2n_\nu}{\ud E_\nu\ud\Omega_\nu}\ud\Omega_\nu\ud E_\nu\\
&=\frac{L_\nu}
{4\pi R_\nu^2\langle E_\nu\rangle}\left[1-\sqrt{1-(R_\nu/r)^2}\right]^2.
\label{eq:ne-eff}
\end{align}
\end{subequations}
We will comment on the validity of the single-angle approximation
at the end of the section.

When they leave the neutrino sphere, 
the neutrinos and antineutrinos form two oppositely directed
NFIS blocks.
For illustrative purposes we assume that $\nu_e$ and $\bar\nu_e$
 dominate the neutrino species emitted from
the proto-neutron star. We will also assume that $\nu_e$ and $\bar\nu_e$
have the same luminosity $L_\nu$, and that
the vacuum coupling coefficients of the two corresponding NFIS blocks are
\begin{equation}
\muV{,1}\simeq\frac{\delta m^2}{2 \langle E_{\nu_e}\rangle} 
\quad\mathrm{and}\quad
\muV{,2}\simeq-\frac{\delta m^2}{2 \langle E_{\bar\nu_e}\rangle},
\label{eq:supernova-muV}
\end{equation}
respectively.

We define a dimensionless quantity
\begin{subequations}
\label{eq:kappa}
\begin{align}
\kappa &\equiv \frac{|\delta m^2|/2\langle E_\nu\rangle}{|\mu_\nu|n_\nu^\eff}\\
&= \frac{|\delta m^2| \pi R_\nu^2}{\sqrt{2}\GF L_\nu}
\left[1-\sqrt{1-(R_\nu/r)^2}\right]^{-2}\\
&\simeq 3.6\times 10^{-6} 
\left(\frac{|\delta m^2|}{3\times 10^{-3}\,\mathrm{eV}^2}\right)
\left(\frac{R_\nu}{10\,\mathrm{km}}\right)^2
\nonumber\\
&\quad\times\left(\frac{10^{51}\,\mathrm{erg/s}}{L_\nu}\right)
\left[1-\sqrt{1-(R_\nu/r)^2}\right]^{-2}.
\end{align}
\end{subequations}
This quantity gives a measure of the inverse of the number
density of either neutrino species.
Using Eqs.~\eqref{eq:bi-polar-cond}, \eqref{eq:ne-eff},
\eqref{eq:supernova-muV} and \eqref{eq:kappa} we find that the 
rough boundary
condition for supernova neutrinos to transition from
the synchronized mode to the bi-polar mode is
\begin{equation}
\kappa \gtrsim \epsilon,
\end{equation}
where the dimensionless quantity
\begin{equation}
\epsilon \equiv 
\frac{(\langle E_{\nu_e}\rangle -\langle E_{\bar\nu_e}\rangle)^2}%
{2(\langle E_{\nu_e}\rangle^2+\langle E_{\bar\nu_e}\rangle^2)}.
\end{equation}
measures the disparity between the energy spectra of $\nu_e$ and $\bar\nu_e$.
If $R_\nu$ is much smaller than 
the boundary radius $r_\mathrm{BS}$ (Bi-polar Starting)
of the two collective modes, we can estimate
\begin{align}
r_\mathrm{BS} &\simeq (51\,\mathrm{km})
\left(\frac{|\delta m^2|}{3\times 10^{-3}\,\mathrm{eV}^2}\right)^{-1/4}
\left(\frac{L_\nu}{10^{51}\,\mathrm{erg/s}}\right)^{1/4}
\nonumber\\
&\quad\times\left(\frac{R_\nu}{10\,\mathrm{km}}\right)^{1/2}
\left(\frac{\epsilon}{0.01}\right)^{1/4}.
\end{align}

As neutrinos propagate away from the proto-neutron star,
the local neutrino density decreases. Beyond some radius
the neutrino density is so low that the collectivity of neutrino
flavor transformation breaks down and neutrinos undergo
conventional MSW flavor evolution.
This occurs if 
\begin{equation}
\Delta \muV{1(2)} \gtrsim |\mu_\nu \Sb_{1(2)}|,
\label{eq:supernova-MSW-cond}
\end{equation}
where $\Delta \muV{1(2)}$ is the half-width of
the distribution $f_{\nu_e(\bar\nu_e)}(\muV{})$ (see the discussion in 
Sec.~\ref{sec:general-synchronization}).
We estimate that
\begin{equation}
\Delta \muV{1(2)} \simeq 
\frac{|\delta m^2|}{2 \langle E_{\nu_e(\bar\nu_e)}\rangle^2}
\Delta E_{\nu_e(\bar\nu_e)},
\label{eq:delta-muV}
\end{equation}
where $\Delta E_{\nu_e(\bar\nu_e)}$ is the half-width
of the $\nu_e$ ($\bar\nu_e$) energy spectra.
Using Eqs.~\eqref{eq:ne-eff}, \eqref{eq:kappa},
\eqref{eq:supernova-MSW-cond} and \eqref{eq:delta-muV}
we can obtain a condition for where collectivity of
neutrino flavor oscillations will break down:
\begin{equation}
\kappa \gtrsim \frac{\langle E_\nu\rangle}{2\Delta E_\nu}.
\end{equation}
If $R_\nu$ is much smaller than
the boundary radius where collectivity breaks down, 
$r_\mathrm{BE}$ (Bi-polar Ending), we can show that
\begin{align}
r_\mathrm{BE} &\simeq (193\,\mathrm{km})
\left(\frac{|\delta m^2|}{3\times 10^{-3}\,\mathrm{eV}^2}\right)^{-1/4}
\left(\frac{L_\nu}{10^{51}\,\mathrm{erg/s}}\right)^{1/4}
\nonumber\\
&\quad\times\left(\frac{R_\nu}{10\,\mathrm{km}}\right)^{1/2}
\left(\frac{\Delta E_\nu/\langle E_\nu\rangle}{0.25}\right)^{-1/4}.
\end{align}

\begin{figure}
\begin{center}
\includegraphics*[width=\myfigwid, keepaspectratio]{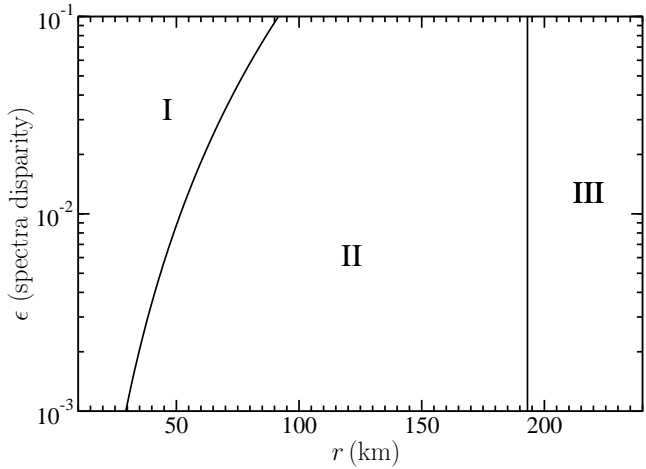}
\end{center}
\caption{\label{fig:boundary-radii}%
The regions where the neutrino oscillations in the supernova environment
are dominated by synchronized (I), 
bi-polar (II)
and conventional MSW (III) flavor evolution. 
In the calculations for this figure we have taken 
$|\delta m^2|=3\times 10^{-3}\,\mathrm{eV}^2$, $R_\nu=10\,\mathrm{km}$,
$L_\nu=10^{51}\,\mathrm{erg/s}$ and $\Delta E_\nu/\langle E_\nu\rangle=0.25$.}
\end{figure}

Taking $|\delta m^2|=3\times 10^{-3}\,\mathrm{eV}^2$, $R_\nu=10\,\mathrm{km}$,
$L_\nu=10^{51}\,\mathrm{erg/s}$ and $\Delta E_\nu/\langle E_\nu\rangle=1/4$, 
we can calculate the
boundary radius between the two collective modes and 
the radius of the boundary separating the collective
modes from the regime where conventional MSW evolution dominates.
These boundaries are shown in Fig.~\ref{fig:boundary-radii}.
It is clear that supernova neutrinos 
are in the synchronized mode near the proto-neutron star
(region I in Fig.~\ref{fig:boundary-radii}),
but could experience bi-polar flavor transformation at a
moderate distance  (region II). It is only in the region far from
the proto-neutron star that  neutrinos will undergo
conventional MSW transformation (region III). This is very different from
the solar neutrino oscillation case, where neutrinos experience only
MSW flavor transformation. We note that there are actually no
sharp boundaries between these flavor transformation regions.
There will be neutrinos and antineutrinos with many different
energies in any region above the proto-neutron star.
As a result, a particular region in general could host superpositions
of various neutrino and antineutrino oscillation modes.
Therefore, regions I, II and III should be
understood as where synchronized, bi-polar, and conventional MSW
type flavor transformations, respectively, \textit{dominate}.

In broad brush, the mixing
parameters ($\thetav$ and $\delta m^2$), the neutrino and antineutrino
energy spectra and luminosities, and $r$ and $R_\nu$ are the
principal determinants of the dominant oscillation mode at
a particular location. Of course, the actual detailed
form of flavor oscillation at any point  is also affected by
the matter density and electron fraction.
For example, if $\nu_e$ number flux dominates
over that for $\bar\nu_e$ and $\delta m^2>0$,
 neutrinos
and antineutrinos in the synchronized mode 
will evolve as if they were one neutrino
with energy $E_\mathrm{sync}$ [Eq.~\eqref{eq:Esync}].
As a result, neutrinos and antineutrinos in this case will experience
flavor conversion at radius $\sim r_\mathrm{MSW}(E_\mathrm{sync})$, where
$r_\mathrm{MSW}(E_\nu)$ is the MSW resonance radius for a neutrino
with energy $E_\nu$. This radius is determined by the standard
MSW resonance condition,
\begin{equation}
\frac{\delta m^2}{2 E_\nu} \cos2\thetav=
\sqrt{2}\GF n_e(r_\mathrm{MSW}(E_\nu)).
\end{equation}

Governed by the density run in the supernova envelope,
neutrinos and antineutrinos may experience the following 
collective flavor mixing scenarios:
\begin{itemize}
\item If the proto-neutron star has a very ``thick'' envelope,
$n_e$ is very large throughout regions I and II and
$r_\mathrm{MSW}(E_\mathrm{sync})\gtrsim r_\mathrm{BE}$. In this case
no significant flavor conversion will occur when neutrinos are
in region I. For $\delta m^2>0$ flavor
conversion is also suppressed in region II. For $\delta m^2 <0$, 
however, large flavor mixing
can occur  in region II.
\item If the envelope of the proto-neutron star is ``thin'',
$n_e$ is very small in regions II and III, and
$r_\mathrm{MSW}(E_\mathrm{sync})\lesssim r_\mathrm{BS}$. 
For $\delta m^2 >0$, almost complete flavor conversion 
($\nu_e\rightarrow\nu_\tau$ and $\bar\nu_e\rightarrow\bar\nu_\tau$)
occurs around radius $r_\mathrm{MSW}(E_\mathrm{sync})$ in region I.
Entering region II, neutrinos and antineutrinos are now dominantly
$\nu_\tau$ and $\bar\nu_\tau$, and large flavor mixing will occur
 (see Table \ref{tab:bi-polar-large-mixing}).
For $\delta m^2 <0$, flavor conversion is suppressed in
region I, and large flavor mixing can occur in region II.
\item If the proto-neutron star has an envelope of a moderate thickness,
$n_e$ is large in region I and part of region II and
$r_\mathrm{BS}\lesssim r_\mathrm{MSW}(E_\mathrm{sync})\lesssim r_\mathrm{BE}$.
In this case, flavor mixing is always suppressed in region I.
For $\delta m^2 <0$, large flavor mixing can occur
 in region II. The exact flavor oscillation form is not clear for
the $\delta m^2>0$ case in region II. However,  some
resonance-like behavior around radius $r_\mathrm{MSW}(E_\mathrm{sync})$
could be expected.
\end{itemize}

The typical collective oscillations described above
are idealized. In particular, we have assumed
that the gradient of $n_e$ is not so large that
 the adiabatic condition is violated. If the adiabatic condition is
violated, other interesting phenomena may occur.
For example, the NFIS's of neutrinos and antineutrinos
may be kicked into a configuration where they are aligned or anti-aligned
with $\basef{x}$ at some instant. This corresponds
to a maximally mixed state. If this occurs in region I
and the adiabatic condition holds from this point on, 
 the NFIS's will rotate collectively around 
$\Hb^\eff=\langle\muV{}\rangle\HV+\HA$.
This nearly maximal  mixing will last until
radius $r_\mathrm{BS}$ or $r_\mathrm{MSW}(E_\mathrm{sync})$,
 whichever is smaller.
This is the Background Dominant Solution described in Ref.~\cite{Fuller:2005ae}.

Our analysis of collective flavor transformation 
assumed isotropy of the neutrino gases. This  is appropriate
for the early universe scenario. In the supernova environment
the neutrino gas is not isotropic. The anisotropy of the supernova
neutrinos has two major effects. One effect is that
the neutrinos scattering at some particular point 
have travelled different distances
from the neutrino sphere before they interact. 
We note that the propagation distances along various
trajectories are most different near the neutrino sphere. 
Close to the proto-neutron star,
the electron density is so large that $n_e\gg n_\nu^\eff$, and
$\HA$ breaks the correlation of the NFIS's on different trajectories.
As a result, the NFIS's on different trajectories  develop
relative phases.
This effect, however, does not compromise our analysis because
neutrinos and antineutrinos are essentially kept in their
flavor eigenstates by $\HA$, and the effects of  destructive
interference are small.
After neutrinos propagate away from the proto-neutron
star, the distance difference between any two trajectories becomes
small.

The other effect of the anisotropy of supernova neutrinos 
is that the neutrino-neutrino forward scattering potential,
and therefore $\mu_{ij}$, depends on the angle between the
directions of the neutrino momenta [Eq.~\eqref{eq:mu-ij}].
As a result, the effective total NFIS 
\begin{equation}
\Sb^\eff_i\equiv\mu_\nu^{-1}\sum_j\mu_{ij}n_{\nu,j}\sB_j
\end{equation}
is different for NFIS's on different trajectories, and
one cannot define a universal total NFIS $\Sb$ in the
original isotropic sense. However, we note that the NFIS's on
different trajectories are still strongly coupled as a result of
 large neutrino density. This is why
collective neutrino flavor oscillations can arise in the first place.
Although the exact neutrino oscillation
behavior can only be shown by the numerical simulations which treat
the trajectory issue self-consistently, we expect that
qualitatively similar flavor oscillations, \textit{e.g.},
large flavor mixing in the $\delta m^2<0$ and $\thetav\ll 1$ scenario, 
may occur in the real supernova environment.

\section{Conclusions\label{sec:conclusions}}

We have introduced a notation for neutrino flavor isospin
which explicitly exhibits symmetry between flavor
transformation of neutrinos and antineutrinos.
 We have pointed out a key quantity in dense gases of
neutrinos and/or antineutrinos, the total effective energy, which is
conserved in some interesting cases. Using the conservation
of the total effective energy, we have  proved
the stability of  synchronized flavor
transformation in a simple and intuitive fashion. We have
also demonstrated how co-rotating frames can be
useful in analyzing collective oscillation in more general cases.

With the concept of total effective energy we have
for the first time explained why large flavor mixing
occurs for a dense gas of initially  pure $\nu_e$ and $\bar\nu_e$ 
with a small mixing angle
and an inverted mass hierarchy. We have estimated
the oscillation periods of  bi-polar systems using simple
dimensional analysis. Additionally, we have studied
more complicated and more general
 bi-polar systems by using
co-rotating frames. We have also for the first time
demonstrated that 
 a dense gas initially consisting of pure 
$\nu_e$ and $\bar\nu_e$  with 
 an inverted mass hierarchy can develop large
flavor mixing, even in the
presence of a dominant matter background.

We have derived a convenient criterion for determining whether
the synchronized or bi-polar type of collective oscillations
may arise in a dense neutrino and/or antineutrino gas.
Based on this criterion, we have estimated the regions
where various modes of flavor oscillation may occur
in the supernova environment.
We have found that neutrinos emitted from the proto-neutron
star in a core-collapse supernova event generally experience
synchronized and bi-polar flavor transformations in sequence
before the conventional MSW flavor transformation takes over.
We have also described the typical flavor oscillation behaviors
according to different density runs in the supernova envelope.

Although our analysis of neutrino flavor transformation 
in the supernova environment
is based on crude estimates, it does
suggest a picture of neutrino flavor transformations
dramatically different from that in the solar case.
In particular, because of the
large neutrino luminosities, both synchronized and bi-polar types
of collective flavor transformations are
involved in the supernova scenario. 

To go beyond this work we would need to drop a number of
the approximations made here and go over to a detailed numerical
simulation starting from realistic conditions of neutrino and
antineutrino luminosity and spectral distribution. Chief
among the requirements of a detailed numerical model
would be a self-consistent treatment of flavor evolution
on different trajectories from the neutrino sphere. This could
be especially important for regions near the neutrino sphere.
Even when such detailed numerical simulations are accomplished,
the collective behavior of neutrino flavors will remain
a complicated phenomenon. This is where our simple physical pictures may be
most useful: delineating the expected qualitative behavior of
the self-interacting neutrino system in various supernova conditions.

In any case, our results have probably overturned some of
the existing paradigms related to the supernova neutrino flavor 
oscillation problem.
Among these, the analyses of future
supernova neutrino signals are certainly
affected, because most if not all
 current analyses are based on the assumption that
the conventional MSW transformation is valid throughout the
supernova environment. Depending how deep the 
collective large-scale flavor mixing
of neutrinos and antineutrinos may occur,
the treatment of 
shock re-heating  and nucleosynthesis might also be affected. 

\begin{acknowledgments}
This work was supported in part by UC/LANL CARE grant,
NSF grant PHY-04-00359,
the TSI collaboration's DOE SciDAC grant at UCSD, and
DOE grant DE-FG02-87ER40328 at UMN. 
\end{acknowledgments}

\bibliography{ref}

\end{document}